# Live-cell imaging powered by computation

Hari Shroff [1], Ilaria Testa [2], Florian Jug [3], Suliana Manley [4]

[1] Janelia Research Campus, Howard Hughes Medical Institute (HHMI), Ashburn, VA, USA

[2] Department of Applied Physics and Science for Life Laboratory, KTH Royal Institute of Technology, 100 44, Stockholm, Sweden

[3] Fondazione Human Technopole (HT), Viale Rita Levi-Montalcini 1, 20157 Milan, Italy

[4] Institute of Physics, School of Basic Sciences, Swiss Federal Institute of Technology Lausanne (EPFL), Lausanne, Switzerland

Corresponding author: suliana.manley@epfl.ch

## Abstract

The proliferation of microscopy methods for live-cell imaging offers many new possibilities for users but can also be challenging to navigate. We focus here on computational methods that promise to boost live-cell fluorescence microscopy, where intra-cellular dynamics and cell viability constrain measurements considerably. Considering the tradeoffs between signal-to-noise ratio (SNR), spatial resolution, temporal resolution, and multi-color and multi-channel capacity, we review computational methods that can be layered on top of commonly used existing microscopies, as well as hybrid methods that integrate computation and microscope hardware.

## Introduction

An essential feature of living systems is that they are dynamic. Molecules form complexes at specific locations and times, initiating and propagating cascades of reactions. Organelles change size, shape, composition, and organization within the cell, exchanging signals and molecules as they contact each other. Cells migrate, divide, and differentiate to specialize within a tissue. Thus, sequences of events encode causal relationships and reveal mechanistic details, from molecular to organismal scales. Fluorescence microscopy, which enables dynamic imaging of specific molecules, their interactions, and biochemical states in living samples, is thus an important tool for biological discovery. Yet, interpreting image data requires quantitative analysis – this is where computation, through automated image processing pipelines, typically takes over. Such analyses have yielded insights into modes and rates of motion, through segmentation and object tracking [1-4]; or molecular exchange and binding kinetics, through photobleaching or photoactivation recovery and single particle imaging [5-8]. However, advances in computational approaches, particularly through deep learning [9-11] (**Box 1**), also have important implications for experimental design, and are changing the boundaries of what is possible to capture in living systems. To better understand the potential of computation, it is useful to consider current limitations to live microscopic acquisitions.

The ideal microscopy measurement of a living specimen captures dynamics with molecular specificity, at relevant structural scales without introducing perturbations. What does this mean practically, in terms of image quality? The signal-to-noise ratio should be high enough to enable quantitative analysis. The spatial resolution should be high enough to distinguish functional molecular organization. The temporal resolution should be high enough to capture dynamic intermediate states. And enough channels, to image fluorophores of different colors or lifetimes, should be collected in parallel to reveal molecular or state sequences that define causal relationships. Additional requirements based on the biological system further constrain the measurement. Different measurement aspects, although independently determined by microscope hardware, are entangled due to their common dependence on the amount of fluorescence emitted by the sample (the "photon budget"). We refer to photon budget both because fluorescent molecules can only emit a limited number of photons before irreversibly photobleaching, and because living samples can endure exposure to limited irradiance before suffering from photodamage. The resulting tradeoffs between these measurement properties leads to the "pyramid of frustration," where an improvement along one dimension requires a sacrifice along others, within a given imaging framework [12]. As an example, acquiring images with less light can prolong imaging experiments, but also results in noisy or lower-resolution data.

Computational approaches can work synergistically with other developments to expand the possible accessible measurement properties, as defined by the photon budget. Other technological advances include developing fluorophores with higher photostability to increase the number of photons per molecule [13-15], harnessing fluorophore transitions to enable super-resolution microscopy [16-19], and shaping the excitation, as in light sheet microscopy and multi-photon microscopy [20,21], to reduce out-of-focus exposure and phototoxicity. Unlike these approaches, computation further extends the range of biological phenomena that can be captured using microscopy without having to change the sample or the microscope hardware.

Here we provide an overview of recent developments in the field of computational microscopy, focusing on those which boost aspects important for live fluorescence imaging: signal-to-noise ratio (SNR), spatial resolution, temporal resolution, and multi-color. We review computational tools which can be easily combined with existing microscopies, or used to create hybrid methods that integrate computation and microscope hardware (**Fig. 1a**). We provide a **Glossary** to define technical terms, and summarize the methods reviewed as well as their advantages, requirements, potential artifacts, how they were demonstrated, and link to their code (**Table 1**). Throughout, we distinguish between tools that are at ready-to-use versus proof-of-concept stage and suggest new opportunities ripe for exploration.

## 1. Improving the signal-to-noise ratio

Signal and noise are fundamentally tied to image quality. In the context of fluorescence microscopy, signal is related to the amount of fluorescence emitted by the sample and captured by the detector. Not all fluorescence is desirable: so-called "background" fluorescence can arise due to non-specific staining, out-of-focus planes, scattering, or sample autofluorescence. Also, since fluorescence emission is a stochastic process, signal intensity fluctuates over time, introducing noise. Imaging systems introduce additional noise, such as detector 'read noise' arising from analog-to-digital conversion, and 'dark current' or thermal noise that is present even in the absence of fluorescence. As a fundamental property of the signal and its detection, noise exists even in the absence of background fluorescence. The relevant parameter is the relative proportion of signal and noise, or the signal-to-noise ratio (SNR), since it determines the useful information that can be extracted from images.

SNR has important implications for live fluorescence imaging: in most experiments, the "photon budget" is limiting. During imaging, fluorophores bleach, thereby lowering the SNR. When the SNR approaches 1, useful signal can no longer be discerned from noise and the experiment is effectively over. While higher SNR is always desirable, for live imaging, sample health must also be preserved [12]. Given that higher SNR can be achieved through increased excitation dose, which increases the production rate of damaging reactive oxygen species, the trade-off between SNR and phototoxicity is salient. Also, methods that afford higher spatial resolution generally deplete fluorescence more with each image than alternatives, implying a trade-off between SNR and spatial resolution. Finally, because noise is intrinsic to the imaging process, no two fluorescence images of the same sample are identical. Thus, any fluorescence image is at best an estimate of the underlying sample, and SNR limits data interpretability.

During pilot experiments, microscopists typically attempt to increase signal and decrease background in sample preparation. This is accomplished by selecting fluorophores to optimize photostability and brightness while minimizing spectral overlap with autofluorescence, and dye targeting strategies to optimize specificity and density of labelling while maintaining biological function. Microscopists also tune their measurement parameters, seeking to identify an illumination intensity which results in sufficient SNR to observe their target of interest over a sufficiently long period, and using filters which cut out autofluorescence while keeping signal, and using optical sectioning to remove out-of-focus fluorescence.

Computational approaches offer additional, synergistic ways to boost SNR, through denoising of images after they are collected, or by dynamically tailoring the exposure during a measurement. There are also computational methods to reduce background fluorescence, for example: subtraction of a constant background; or subtraction of local background in the special cases of isolated objects (ImageJ) or single molecules [22]. Here, we review methods focused on SNR enhancement, rather than on background removal.

## 1.1 Classical algorithms for computational denoising

Once sample labeling has been optimized, computational denoising can further improve SNR (**Fig. 1b, 3**). The simplest forms of denoising combine the signal from neighboring pixels, by binning them via a mean or median operation. Binning is fast and freely and easily accessible via open-source software packages such as Fiji [23]. However, it sacrifices spatial resolution in favor of improved SNR; other filtering-based methods, such as median filtering [24], maintain pixel resolution while still reducing noise.

More sophisticated methods that make a priori assumptions about the sample such as total variation regularization [25], or non-local means [26] can be powerful for denoising, but require careful, and often manual, parameter tuning. As one such example, by assuming the sample is spatially continuous within the resolution limit and temporally continuous over multiple frames (in other words, smooth in space and time), Hessian SIM reconstruction [27] allowed 2D super-resolution imaging at roughly one tenth the illumination dose typically required. This improved sensitivity permitted the capture of ~10x more time points than previously possible, enabling the study of highly dynamic endoplasmic reticulum loops and exocytic fusion intermediates, as well as cytoskeletal remodeling on the hour timescale [27].

## 1.2 Deep learning algorithms for computational denoising

Currently, the most potent denoising methods are based on deep learning algorithms [28] (**Box 1, Fig. 1b, 2**), which use neural network architectures. In supervised methods, pairs of low SNR and corresponding high SNR ground truth examples are provided as network training data. Millions of network parameters are adjusted during the training procedure, as the network 'learns' to map from low SNR images to predict the equivalent high SNR images. The trained network can then be applied to previously unseen noisy data, to generate a higher SNR prediction. Since information about the underlying data is incorporated into the network during training, such approaches can quantitatively outperform classical, data-agnostic denoising methods **(Box 2, Fig. 2)**, with important caveats (**Box 3**).

Since the illumination intensity can be lowered 10-100x with deep-learning based denoising [11,29-32], the increased photon budget may be spent in ways that enable fundamentally new measurements. For example, content aware restoration (CARE) [11] restored noisy and otherwise hard to analyze spinning disc confocal microscopy images of nuclei in the planaria *S. meditteranea*, a flatworm with extreme sensitivity to light. It also improved the segmentation of noisy nuclear images in the developing beetle *Tribolium castaneum*. CARE works by training a neural network on pairs of images that represent the same object captured in low-SNR and high-SNR. Whenever such training pairs of images can be acquired, CARE-based image denoising and restoration is applicable.

Denoising of data from more advanced microscope setups such as classical 2D- [33] and 3D-SIM [16], as well as lattice light sheet microscopy-based SIM [34]has also employed deep learning. A different network architecture, *i.e.*, a residual channel attention network (3D RCAN), was designed to denoise fluorescence microscopy volumes [29]. Denoising low-SNR instant SIM (iSIM) [35] recordings of mitochondria in living U2OS cells enabled the collection of 2600 super-resolution volumes without detectable photobleaching or photodamage, improving experiment duration by several orders of magnitude. This imaging duration and gentleness is reminiscent of that attained with light-sheet microscopy, but the denoised iSIM volumes afford higher resolution.

In a further example, knowledge of the physics of image formation was combined with denoising to yield high quality reconstructions [32] with fewer artifacts than previous methods. This method was used to resolve the fast kinetics of beating cilia and intraflagellar transport, the rapid interactions between membrane-bound and membrane-less organelles, and the assembly and disassembly of nucleoli within mitotic cells [32] – all applications that would have been difficult or impossible to observe without powerful denoising methods.

## 1.3 Adaptive illumination to increase SNR

Microscopists often attempt to find the minimum excitation intensity that will achieve a desired SNR, to avoid photobleaching and phototoxicity. However, expression levels and local concentrations of different proteins may differ by orders of magnitude; thus, each structure in a cell or tissue can have a different optimal excitation intensity. Moreover, biological samples are heterogeneous in space and time, implying that illumination patterns designed to achieve a uniform SNR would be non-uniform in space and time. For example, regions with fluorescence below a threshold value – implying no signal of interest -- could be skipped over, while regions with fluorescence above the threshold could be illuminated to maintain a given SNR. Such decisions can be delegated to an automated microscope controller, which may be faster and more accurate than a human. In this manner, coordination between the sample fluorescence and the microscope hardware components through a feedback loop can execute complex acquisitions.

Adaptive illumination strategies based on feedback between the sample and hardware have been demonstrated in different microscopy modalities, from conventional to super-resolution approaches. Controllers for point-scanning confocal [36], two-photon [37,38] , STED microscopes [39], and SIM [40] have been designed to automatically tune the illumination intensity according to the amount of detected fluorescence. This adjustment is typically achieved on the microsecond timescale and can be done point-by-point in a local manner as the illumination is scanned across the sample.

Reducing the sample exposure to focused light is particularly important for STED microscopy, where fluorescent molecules are exposed to the high intensities of the depletion donut, increasing the risk of photodamage. If no photons are detected in a sample location, RescueSTED [39] assumes no fluorescently labelled structures are present, and switches the depletion laser OFF while moving to the next point in the sample. More advanced versions named MINFIELD [41] or DyMIN [42] increase the depletion intensity only when a structure is centered, to provide the maximum spatial resolution on structures of interest. This adjustment minimizes exposure of the fluorescent molecules to the high intensities of the donut crests and allows photobleaching reduction up to 100-fold.

Light exposure in point scanning systems is determined by laser intensity and dwell time. To reduce unnecessary illumination, feedback loops can tell the scanner to illuminate a certain region of the sample until the desired SNR is reached as adopted in Smart RESOLFT [43] and Hopping STED [44] microscopes. This allowed peroxisomes and mitochondria to be imaged at up to 30 Hz in clutered cells, and actin rearrangements to be imaged at a depth of 20–30 µm inside a developing *Caenorhabditis elegans*. Automatic re-adjustment of the illumination in a sample adaptive manner has also been beneficial for single molecule localization microscopy [45], where the photo-activation light was dynamically adapted to achieve efficient recording of sequential single molecule events.

Real-time feedback as described here is an advanced approach, and thus far has only been applied to limited cases and methods where photobleaching is consistently a major concern, such as in super-resolution microscopy. However, feedback is becoming increasingly accessible, as microscopists, core facilities, and commercial developers aim to execute more complex acquisitions [46-51].

# 2. Improving spatial resolution

Blurring introduced by diffraction results in a fundamental information loss when the underlying sample is imaged with any fluorescence microscopy modality. Subcellular structures are frequently of interest, but are most affected by blurring since their dimensions are typically near or below the diffraction limit. Thus, methods that reverse spatial blur-induced resolution loss are of great value for biological imaging. In 2014, the Nobel Prize in Chemistry was given for a family of 'super-resolution' microscopy methods that cleverly circumvent the diffraction limit. These techniques, which differ considerably in practical implementation (**Glossary**) have been reviewed extensively elsewhere [52]. We note that these methods make no assumptions about the sample structure, but have trade-offs -- the greater the resolution gain, the higher the cost in SNR, phototoxicity, or temporal resolution -- making some better suited for live imaging than others [53].

In this section, we describe computational strategies that alleviate these tradeoffs; developments in deconvolution methods that enhance the resolution and contrast of the raw data; difficulties that arise when attempting to maintain resolution in aberrating tissue; and real-time feedback methods for achieving improvements in spatial resolution without sacrificing SNR or sample health.

## 2.1 Deconvolution

The blurring introduced by fluorescence microscopy can be considered analogous to a low pass filter, in that the microscope transmits all spatial frequencies up to the resolution limit. Low spatial frequencies (corresponding to larger features in the sample) are transmitted at higher SNR than high spatial frequencies, leading to a resolution-dependent loss in contrast that can obscure fine details in low-resolution haze, particularly for 3D samples. Fortunately, if the blurring function and noise characteristics of the microscope are well characterized, this degradation can be partially reversed using a computational procedure known as deconvolution [54] (**Fig. 3 a-c**). The main practical benefit is a large improvement in contrast and SNR, which in turn leads to discernible improvements in spatial resolution over the raw input data.

The deconvolution literature is vast, and properly surveying it would take its own review. Where should a new user begin? Classical deconvolution algorithms are still effective and remain broadly used. For example, the decades-old Richardson-Lucy [55,56] (RL) algorithm iteratively deblurs an estimate of the sample given a known PSF and assumed Poisson noise (which is usually dominant in fluorescence microscopy) [57].

Deconvolution methods that incorporate additional information about the specimen can considerably improve upon RL, in some cases restoring fine detail outside the classical resolution limit. This is accomplished either by assuming an underlying smoothness [27] or sparsity [58] to the sample, or by explicitly incorporating sample information into deconvolution via deep learning [59]. Such improved performance comes at a cost, requiring either careful parameter tuning [27,58] or training data [59].

Finally, deconvolution can be used beyond its 'traditional' context in deblurring images captured with conventional fluorescence microscopes. Deconvolution is often at the heart of 'computational microscopes', including multi-view light-sheet [60,61] or confocal [62] imaging systems that improve volumetric resolution and SNR by extracting the best information from each view. Deconvolution can also be used to combine information from different microscopes [63] -- enabling, for example -- kHz activity imaging of acetylcholine and voltage deep in mouse brain with rapid line-scanned multiphoton tomography after first acquiring a static high-resolution volume with traditional raster-scanned multiphoton microscopy [64].

## 2.2 Deep learning-based resolution enhancement

By contrast to the 'classical' super-resolution techniques mentioned above, resolution enhancement can also be enabled by embedding prior knowledge about a sample's structure into a trained neural network, with some important caveats **(Box 4)**. By acquiring matched low- and high- resolution training data, such 'cross modality' resolution enhancement neural networks have improved the spatial resolution of TIRF [65] and widefield [31] microscopy using SIM ground truth, confocal microscopy using line-scanning SIM [62] and STED microscopy ground truth [29,65], and instant SIM using expansion microscopy ground truth [29]. Although none of these methods perfectly reproduces the higher resolution modality, the main benefit is that resolution is enhanced while retaining the higher acquisition speed, SNR, and/or phototoxicity advantages of the lower resolution method (**Fig. 3 d,e**). Even greater benefits accrue when using multi-step networks trained to sequentially denoise and enhance resolution [29,62,66]. Collectively, these methods have revealed mitochondrial nucleoid dynamics in the context of mitochondrial cristae and the rotational streaming of mitochondrial tubes [31]; mitochondrial fission [30], cell division [29], and cytoskeletal dynamics within immune cells [29,66]; and subcellular growth cone dynamics in living *C. elegans* embryos [62]. The inferred high-resolution images can enable more accurate segmentation and reveal finer detail than lower resolution raw data.

Deep learning can also compensate for the anisotropic PSF present in most optical microscopes [11,67], which results in two- to three-fold worse resolution axially along the optical axis than laterally in the imaging plane. The concept is to create isotropic image volumes by blurring and downsampling image stacks in the lateral dimension, so that lateral views resemble the axial data recorded by the microscope. A neural network can reverse this degradation to improve resolution in the axial views, given the high-resolution lateral ground truth from the original images, and the assumption that structures look similar in both axial and lateral views. This method has been applied to volumetric data from fruit fly, zebrafish, and mouse liver, in the former case increasing resolution sufficiently to quantitatively improve nuclear segmentation [11]. Similar concepts have been used to improve isotropy in multiview confocal super-resolution microscopy [62], revealing dynamic histone nanodomains in live Jurkat T cells and offering clear views of the nerve ring region in anesthetized larval *C. elegans*; and in 3D SIM, enabling high contrast volumetric time-lapse imaging of organelle dynamics at 120 nm isotropic resolution [66].

## 2.3 Adaptive Optics to correct spatial aberrations

High- or super-resolution imaging is much easier to achieve in single cells than in tissue, in part because thicker samples contain more interfaces and sub-regions which can bend light. Such bending can significantly distort the wavefront of light, causing optical aberrations that prevent diffraction-limited focusing, reducing spatial resolution, contrast, and SNR. Given the importance of studying living cells away from the coverslip in a more physiological context, adaptive optics (AO) methods [68,69] that can measure and subsequently correct the aberrated wavefront are of great value.

The key challenge in AO is determining the aberrated wavefront; once known, the aberration can be corrected by a deformable mirror or spatial light modulator, programmed to apply an equal and opposite aberration. Methods can be broadly classified as 'direct' or 'indirect.' Direct methods introduce point-sources with known emission profile and use dedicated hardware to sense deviations from the expected diffraction-limited emission. Indirect methods iteratively estimate the wavefront by applying known aberrations while monitoring an image quality metric. Direct wavefront sensing tends to be faster, while indirect methods are better suited to imaging in opaque tissue; both have been applied to substantially improve image quality in live imaging studies conducted with confocal [70], light-sheet [71], multiphoton [72,73], and structured illumination [74,75] microscopy.

Despite these successes, existing AO implementations also introduce more illumination dose, degrade temporal resolution, and add substantial complexity and cost relative to the base microscope they are designed to enhance. These issues have hindered widespread adoption of AO and have confined it to the province of relatively few labs. Improving the performance and accessibility of AO is an important frontier in computational microscopy and could have an outsized impact given that the vast number of microscopes suffer from increasing depth-dependent aberrations. Although existing efforts are nascent [76-78], we suspect that deep learning could have an important part to play in the future.

## 2.4 Real-time feedback and inference of sample properties to enhance spatial resolution

Feedback loops that link sample-based computation to the microscopy hardware for adaptive acquisitions can be used to enhance spatial resolution. One such example is MINFLUX microscopy [79], which reduces the number of photons required to localize single molecules. This is enabled by sequential, patterned illumination of isolated fluorophores, coupled with real-time computation to infer their positions. In response, the microscope re-positions the illumination pattern within microseconds according to the estimated emitter location, thereby localizing the emitter to a few nm thus outperforming the standard method of uniform illumination followed by centroid-finding with orders of magnitude fewer photons. With this efficient use of the photon budget, the individual steps and rotations of single motor proteins could be tracked in three dimensions in vitro and in cells, at physiological ATP concentrations [80,81]. MINFLUX has the potential to make localization microscopy more compatible with time-lapse imaging, although bottlenecks remain related to the need for high-density labeling to resolve structures, the label size which often exceeds the nanometric localization precision, and the difficulty of isolating individual fluorophores.

Light-sheet microscopy, while among the gentlest microscopy approaches, can also make use of feedback-loops to increase image quality. The AutoPilot [82] framework dynamically adapts properties such as light sheet intensity and position to maximize focus and contrast in real-time, even as the imaged embryo undergoes major changes in size, fluorophore expression, concentration and positioning during organogenesis.

Artificial intelligence can enhance the performance of data-driven microscopy control by aiding decision making in complex settings. In learned adaptive multiphoton illumination [83] (LAMI), a trained neural network selects intensities while imaging 3D scattering samples such as lymph nodes. Here the choice of a specific laser intensity is based on a complex relationship between fluorescence intensity, incident excitation power and sample shape. Acquisition parameter choice was also guided by machine learning for STED and confocal imaging, to enhance image quality by optimizing laser power and pixel dwell time [84].

# 3. Improving temporal resolution

The temporal resolution of a time-lapse microscopy experiment is defined as the amount of time between images. It is also characterized by the imaging speed or frame rate, which is typically reported as the number of frames per second (fps), or equivalently Hertz (Hz, 1/s). In microscopy, the minimum value of the temporal resolution is the time it takes to acquire a single image. For dynamic processes taking place on timescales that are slow compared to the time to acquire a single image, a delay between frames may be added, which is then also added to the temporal resolution. Without sufficient temporal resolution, biological processes can appear distorted by motion blur if objects move during the capture of a single frame, or can be missed altogether. A valuable concept for thinking about the measurement parameters needed to resolve a process is the Nyquist-Shannon sampling theorem, which defines the minimum spatial or temporal sampling rate required.

Time resolution is sample limited: proteins or other biomolecules of interest may be present in the cell at low abundance; thus, the sample must be exposed to enough light to achieve the desired SNR. This means that exposure times must be long enough, or, in the case of focused excitation, dwell times must be sufficient to gather enough photons at each location in the sample, given that the fluorescence lifetime is typically a few nanoseconds. As a result, denoising (**Section 1**) can have the added value of reducing acquisition time, since noisier data can be collected more rapidly.

Hardware does not usually limit the time resolution when imaging two-dimensional samples -- for widefield illumination, detectors are cameras, and high-end cameras such as sCMOS (scientific complementary metal oxide semiconductor) sensors can achieve frame rates greater than one thousand fps, much faster than most ~~visible~~ cellular processes. When using focused illumination, imaging speed depends on the time to scan the excitation pattern over the sample while maintaining sufficient SNR at each location. The detector is typically a photomultiplier with nanosecond response times; thus, for single point measurements such as in fluorescence correlation spectroscopy of molecular diffusion, the time resolution is more than adequate. However, the acquisition time scales with the scanned sample area, and is multiplied by the number of planes in volumetric imaging, which can often present a practical bottleneck that limits acquisition speed.

## 3.1 Parallelized scanning microscopes

Parallelization collects information from multiple sample areas simultaneously. Thus, it offers a means of increasing imaging speed by reducing the scan range needed to image a field of view. One approach for imaging rapidly in a single plane is to use multi-focal excitation, such as in spinning disk confocal microscopy. Compared with point scanning, the speed to record a field of view increases proportionally to the total number of foci in the disk. The emission from the foci is collected by a camera, allowing acquisition rates of several hundred fps. The main tradeoff compared with a point scanner is lateral and axial crosstalk, which can degrade SNR and optical sectioning in thick or densely labeled samples.

Several super-resolution techniques have taken a similar approach, boosting resolution through computational post-processing, or optical image processing. Lateral resolution in confocal microscopy can be improved by detecting the fluorescence emission with a camera and reassigning the light to account for blurring due to off-axis detection [85]. A similar gain in resolution can be achieved by closing down the pinhole of the confocal microscope, but image reassignment offers higher SNR compared to this approach, since light that would otherwise be rejected by the pinhole can be collected and used. This reassignment [86] is available in some commercial microscopes (e.g., the Zeiss Airyscan). A parallelized version (multifocal SIM, MSIM) also has been invented, using an array of excitation beamlets[87,88]. This enabled imaging cytoskeletal microtubules and actin in 3D cultured cells and living zebrafish embryos at ~1 fps.

In Re-Scan Confocal [89] and Optical Photon Reassignment [90] microscopies, optical image processing performs the re-assignment in a single camera exposure. This concept has also been parallelized in the instant SIM (iSIM) [35] and related solutions [91], which take advantage of thousands of parallel excitation beamlets and matched detector pixels to achieve speeds of ~100 fps, with three-dimensional optical sectioning performed by a pinhole array. Commercial solutions are available, and standard fluorophores can be used, with photostability a desirable quality. These methods all offer up to a two-fold improvement in spatial resolution, derived from photon reassignment and deconvolution (Section 2.1).

The super-resolution method RESOLFT [92] is also compatible with live-cell imaging and parallelization. RESOLFT microscopes are challenging to construct because they rely on patterned illumination to photo-switch the fluorescence, so illumination patterns must be carefully designed and aligned. As for every scanning microscope, the time to collect an image increases with the number of scan positions required i.e., with the size of the field of view. Parallelization has also been used to speed up and expand the field of view RESOLFT, by creating an array of more than 100,000 nanosized regions in the sample recorded at the same time [93]. This allowed intermediate filaments and cytoskeletal actin to be imaged in living cells at 2 fps, with resolutions below 100 nm. These nanosized regions can also be confined axially to enable 3D super resolution imaging in living cells by switching the sample with periodic illumination patterns that are also modulated in the axial direction [94]. With this approach, actin filaments and mitochondrial networks were resolved volumetrically, and their spatial organization quantified in living cells.

## 3.2 Deep-learning approaches to increase imaging speed

Many super-resolution methods are slower than diffraction-limited methods due to the time and light dose required to photoswitch fluorophores. Deep learning approaches can be used to increase the speed of data acquisition for these methods.

In some methods, a neural network attempts to directly predict a higher resolution image from a single low resolution input image, benefiting from the higher imaging speeds of lower resolution methods (Section 3.2). Deep learning can also speed up acquisitions for single-molecule localization microscopy (SMLM) [95]. SMLM pools sparse molecular localizations gathered from many diffraction-limited images to generate a super-resolution image. Collecting sufficient localizations to sample the structures of interest takes hundreds or thousands of raw images, generally limiting the method to fixed cells. Relaxing the requirements for single molecule isolation speeds up sampling but makes precise localization more challenging. Because single molecule images are relatively simple to model, synthetic data can provide a large amount of training data, in the absence of ground truth (molecular positions) for real data. The DECODE [96] network trained on synthetic data could predict molecule locations from high-density data, where multiple emitters were turned on simultaneously. This procedure detected more localizations with better precision than all previous approaches, while simultaneously lowering the number of raw data frames required to reconstruct an image. These advances improved the speed and lowered the phototoxicity of SMLM data collection (70% less illumination and 7x faster than previously possible), thereby enabling the dynamics of the Golgi apparatus and endoplasmic reticulum to be reconstructed from 7.5 s of raw data, and nuclear pore complexes from 3 s of raw data.

### 3.3 Sample-adaptive imaging speed

Dynamic changes in biological samples can be used to guide microscopy experiments without user intervention, and prioritize using the limited photon budget to image rapidly on demand. This is especially valuable in the case of rapid or transient biological events.

For example, mitochondrial fissions occur intermittently every few minutes, and proceed over a few seconds. When observed with techniques such as iSIM [35], which provide higher spatial resolution compared to conventional microscopy, photobleaching and phototoxicity limit the duration of the observation. In event driven acquisitions (EDA) [97], a neural network was trained to recognize mitochondrial precursors to fission. The network was incorporated into the microscope control feedback loop, to trigger faster imaging during fission. This improved the chance of recording structural intermediates during fission.

Calcium concentration or pH changes occur rapidly, on the millisecond timescale, during signaling processes. Such changes can be detected by fluorescent sensors, which will display spikes or gradual increases in fluorescence signal. Exo- or endo-cytosis generates calcium spikes, which were detected with widefield microscopy at a high imaging speed (5-10 ms) throughout a large sample region and used in event-triggered STED (etSTED) [98]. The appearance of a spike triggered a shift in imaging modality, from low to high spatial resolution: STED imaging was initiated only in the sub-cellular region where the calcium spike was detected, saving both time and the rest of the sample from high intensity exposure.

## 4. Multicolor Imaging

Visualizing multiple interacting cellular or tissue components requires tagging them with distinguishable fluorescent markers, or fluorophores. Such markers can either be fluorescent dyes or fluorescent proteins. The engineering and subcellular targeting of fluorescent dyes has a long history, and soon after the first fluorescent proteins (FPs) had been cloned [99], the engineering of FP variants commenced. Today, thousands of dyes and FPs are available for use [100].

Fluorophores selected for concurrent use should be chosen such that their emission spectra overlap as little as possible lest the fluorescence signal in a detection channel intended for one probe will contain contributions from other channels, referred to as “bleed through”. A similar consideration also holds for excitation spectra, where the undesired excitation of multiple structures by a single excitation wavelength is referred to as “crosstalk”. To reduce bleed through and crosstalk, excitation filters or precise laser excitation wavelengths can be used with emission filters to create a dedicated channel for each fluorophore. FPbase [100] is an excellent open and community editable resource for designing experiments, and enables the selection of specific fluorophores in combination with commercially available excitation sources and filters, to reduce bleed through and crosstalk.

Even with dedicated setups that are carefully optimized for a particular fluorophore combination, it is uncommon to see more than two structures being imaged simultaneously in living specimens. Emission wavelengths range between 400-800 nm, and the emission spectral profile for most fluorophores typically span a band between 50-150 nm in width. Thus, on top of increased imaging times and overall light exposure to the sample, with more than three colors the broad emission and excitation spectra make it increasingly difficult to avoid crosstalk and bleed through.

The main computational approach for increasing the number of channels is spectral unmixing. Instead of imaging multiple fluorescent channels sequentially, a spectrum of intensity versus emitted wavelength is recorded for each detection pixel. The spectrum is used to compute the contribution of each fluorophore to each pixel's intensity, thereby estimating the proportion of each fluorophore in each pixel. In contrast to non-spectral imaging, where photons of all wavelengths fall on a single detector surface, a spectral detector typically has multiple separate detection areas that receive different bands of the emitted wavelength, hence recording a discretized spectrum for each pixel. Another set of methods, based on machine learning inference from label-free images, represents a developing area of research [101,102].

## 4.1 Linear unmixing

In the absence of noise, estimating the relative contribution of each fluorescent marker is possible when the number of spectral channels (detected wavelength bands) are at least as many as the number of emitting dyes or FPs. Mathematically, this is equivalent to solving a linear system of $n$ equations (which satisfy the constraint that the sum of all individual emissions at each band is equal to the measured total value) with $n$ unknowns (i.e., total emission per fluorescent marker). This approach is known as "linear unmixing". For linear unmixing to work, the reference spectra of all fluorescent markers contained in the sample are needed. This is achieved by measuring each spectrum individually, or by manually identifying regions in an acquired image that contain only one of the fluorescent markers. Automated component extraction (ACE) procedures exist commercially and can potentially help to avoid this manual calibration step.

In any experiment, the acquired data will be subject to noise, e.g., Poisson shot noise and Gaussian readout noise. The presence of noise introduces uncertainties so that linear unmixing produces multiple solutions. Intensity saturation, sample photobleaching, autofluorescence or other undesired sources of light being imaged (e.g., room lights) also degrade the performance of linear unmixing. In the next sections we will discuss unmixing strategies that better address these issues.

## 4.2 Blind unmixing

Blind unmixing enables fluorophore separation without prior knowledge of emission spectra and a minimal number of detection channels, typically less than the number of fluorophores to be unmixed. One such algorithm, LUMoS (Learning Unsupervised Means of Spectra) [103], learns the relationship between pixels in the raw imaging data and intensity patterns and uses this information to re-classify each pixel into a fluorophore group. This unsupervised approach was applied to two-photon microscopy data and separated up to six channels using only four detectors, which helped to reduce hardware complexity. The accuracy of the separation algorithm was also tested under challenging conditions such as low SNR, significant autofluorescence, and structures of different size.

Another approach that does not require reference spectra measurement is PICASSO (Process of ultra-multiplexed Imaging of biomoleCules viA the unmixing of the Signals of Spectrally Overlapping fluorophores) [104], which enables the separation of up to five spectrally overlapping fluorophores excite with a single laser line and 15 fluorophores in heterogenous brain tissues with the use of multiple excitation laser lines. This approach is based on iteratively minimizing the mutual information between mixed images. If a structure is labelled with two or more fluorophores, the mutual information (MI) of the fluorescence generated in that region is very high. The algorithm tries then to minimize MI by iteratively subtracting scaled images. For this strategy to work, it is necessary to record a number of raw mixed images N equal to the number of fluorophore species, and where one fluorophore species generates the brightest signal in each spectral window.

## 4.3 Unmixing spectral and fluorescence lifetime signals

Linear and blind unmixing strategies can also be applied to fluorescence lifetime imaging microscopy (FLIM) signals, incorporating information from the kinetics of fluorescence emission to identify molecular signatures. Recording FLIM data can be slow, since many photons are needed to reconstruct the fluorescence lifetime decay. Recent developments in electronics and detectors improved the acquisition speed, enabling simultaneous lifetime and spectral detection. Data processing is not trivial due to the size of the raw data, which contains both spectral and temporal information on each pixel of the image.

The Phasor approach [105] was developed to simplify data visualization and reduce post processing time. It provides a graphical overview as a Phasor plot: a two-dimensional histogram of the processes affecting the fluorescence lifetime decay occurring at each pixel. Long and short lifetime decays are located at different positions, and single and multi-exponential decays are also separated. The phasor approach is commercially available and can be used for recording multiple molecular species at the same time, labeled with fluorophores with different spectra or lifetimes. Phasors have also been used to improve the quality of super resolution STED data by efficiently separating the fluorescence signal carrying low resolution information [106].

Blind approaches such as Phasor S-FLIM can unmix images with a sufficiently large number of FLIM photons (tens to hundreds per pixel) [107]. Additional approaches based on deep learning have also been applied to hyperspectral fluorescence lifetime imaging data. UNMIX-ME (unmix multiple emissions) [108] is capable of quantitative fluorophore unmixing by simultaneously using both spectral and temporal signatures. While these algorithms have shown great potential in separating multiple molecular species, they are still dependent on specific microscopes. Future implementations compatible with more general data types will likely lead to increased use.

# 5. Discussion and outlook

We have highlighted recent developments in computational approaches which we find to be most promising in terms of their immediate or near-term potential for improving live-cell imaging. Although we divided techniques according to their improvement to SNR, spatial resolution, temporal resolution, and multi-channel imaging, some of these distinctions are practically convenient rather than fundamental. This is because all axes are linked to photon budget and light exposure, and in that sense gains along one axis can be traded for gains along another. For instance, an acquisition that uses denoising to cope with lower SNR can image faster, or with higher resolution, or in more color channels – or find a compromise and improve all these to a lesser extent. The optimal compromise will depend on the scientific question under investigation.

Computational approaches, especially those based on machine learning, represent a vibrant area of research, whose current limitations are also important to understand [109]. We expect that future developments will improve the specific tools mentioned here. In that regard, it is important to place into context significant improvements that users should strive to adopt, versus marginal gains that could be important in special cases. One may draw parallels here with the development of super-resolution localization microscopy, in which algorithms and software packages proliferated during a flurry of activity. In that case, a valuable resource for the community was a benchmarking challenge that allowed developers and users to more directly compare the performance [57]. A similar resource would be welcome in each of the application spaces described here. In addition, no general-purpose tool currently exists for assessing uncertainty in predictions that a network spits out. Such a tool would be enormously beneficial to a biologist end-user, offering them a 'confidence map' of sorts.

Most ML-based approaches are model-free, which can be an advantage. However, this may also limit their performance -- humans remain better able to generalize and learn from much less data. Centuries of science have relied on model building to “make sense” of patterns, whether physics-based, mathematical, or more heuristic, graphical schema often found in biology. In the case of microscopy, optical physics and engineering are reliably used to quantitatively model and interpret biological imaging experiments. These priors could be used to validate and further improve existing neural network methods, or reduce artifacts.

Multicolor imaging and imaging at depth are areas that we think will see exciting growth in the coming years. It is still difficult to routinely image with more than a few colors, and it is still hard to image deep inside samples with the clarity that we can obtain when imaging at the coverslip. Given the difficulties associated with pure hardware solutions, we suspect computation will have an important part to play in these efforts. So far, purely computational methods are less demonstrated to go directly from low temporal resolution to high temporal resolution or small imaging volumes to large ones. These contexts are especially challenging for ML approaches, since ground truth data is difficult to obtain [110]. Moreover, many dynamic processes in biology are intermittent, nondeterministic, and take place across multiple timescales – thus, interpolation is unlikely to reliably capture them. Similarly, for volumetric imaging inferring content in unseen planes assumes continuity, but structures may be discontinuous and contain information across multiple length scales. Thus, we expect that combining computation and hardware will offer a promising approach to these challenges.

We anticipate that integration between computation and microscope hardware will continue to develop, to enable more informative automated data acquisitions. Early examples of smart microscopy have already demonstrated how biological context can be used to adapt acquisitions. By prioritizing SNR, temporal and spatial resolution, or multicolor imaging for events of interest or optimizing image quality in real time, it is possible to more efficiently capture data relevant to test hypotheses. This offers a way to repurpose ML networks developed for segmentation or other image processing tasks, to improve the data at its source. Extending beyond photon budget limitations, the fields of systems biology and neuroscience have made significant inroads into using optogenetic control to dissect genetic [111] or neuronal [112] circuits using real-time activation and feedback loops. These methods, combined, may eventually allow concepts used to derive natural physical laws algorithmically from data [113] to be extended to biological systems, by determining which experiments should be done next to best constrain a model.

The complexity of an end-to-end microscopy experiment has increased enormously in recent years, as biological samples are prepared to better resemble native specimen, microscopes leverage advances in optics and automation, and analyses attempt to recover as much information as possible. This can present daunting challenges for researchers who lack access to advanced infrastructures. A dream solution would be enhanced collaboration between microscope users, and hardware and software developers. As examples, Janelia's Advanced Imaging Center and the Eurobioimaging infrastructure offer access to imaging technologies, training and support. While these efforts have outsized impact by allowing access to a broad range of users independent of their institutional affiliations, more is needed to encourage exchanges in the area of computation [114] : for example, through consortia like AI4Life, as well as resources like Micromanager [115] and the BioImage Model Zoo [116].

# References

1. Ducret, A., Quardokus, E.M. & Brun, Y.V. MicrobeJ, a tool for high throughput bacterial cell detection and quantitative analysis. *Nat. Microbiol.* **1**(2016).
2. Levet, F. et al. SR-Tesseler: A method to segment and quantify localization-based super-resolution microscopy data. *Nat. Methods* **12**, 1065-1071 (2015).
3. Rizk, A. et al. Segmentation and quantification of subcellular structures in fluorescence microscopy images using Squassh. *Nat. Protoc.* **9**, 586-596 (2014).
4. Ulman, V. et al. An objective comparison of cell-tracking algorithms. *Nat. Methods* **14**, 1141-1152 (2017).
5. Axelrod, D., Koppel, D.E., Schlessinger, J., Elson, E. & Webb, W.W. Mobility measurement by analysis of fluorescence photobleaching recovery kinetics. *Biophys. J.* **16**, 1055-1069 (1976).
6. Elson, E.L. & Magde, D. Fluorescence correlation spectroscopy. I. Conceptual basis and theory. *Biopolymers* **13**, 1-27 (1974).
7. Gahlmann, A. & Moerner, W.E. Exploring bacterial cell biology with single-molecule tracking and super-resolution imaging. *Nat. Rev. Microbiol.* **12**, 9-22. doi: 10.1038/nrmicro3154. (2014).

8. Saxton, M.J. & Jacobson, K. Single-particle tracking: Applications to membrane dynamics. *Annu. Rev. Biophys. Biomol. Struct.* **26**, 373-399 (1997).
9. Moen, E. et al. Deep learning for cellular image analysis. *Nat. Methods* **16**, 1233-1246 (2019).
10. Van Valen, D.A. et al. Deep Learning Automates the Quantitative Analysis of Individual Cells in Live-Cell Imaging Experiments. *PLoS Comput. Biol.* **12**(2016).
11. Weigert, M. et al. Content-aware image restoration: pushing the limits of fluorescence microscopy. *Nature Methods* **15**, 1090-1097 (2018).
12. Laissue, P.P., Alghamdi, R.A., Tomancak, P., Reynaud, E.G. & Shroff, H. Assessing phototoxicity in live fluorescence imaging. *Nat. Methods* **14**, 657-661 (2017).
13. Grimm, J.B. et al. A general method to improve fluorophores for live-cell and single-molecule microscopy. *Nat. Methods* **12**, 244-250 (2015).
14. Lukinavičius, G. et al. A near-infrared fluorophore for live-cell super-resolution microscopy of cellular proteins. *Nat. Chem.* **5**, 132-139 (2013).
15. Shaner, N.C. et al. Improving the photostability of bright monomeric orange and red fluorescent proteins. *Nat. Methods* **5**, 545-551 (2008).
16. Gustafsson, M.G.L. et al. Three-dimensional resolution doubling in wide-field fluorescence microscopy by structured illumination. *Biophys. J.* **94**, 4957-4970 (2008).
17. Hell, S.W. Far-field optical nanoscopy. *Science* **316**, 1153-8 (2007).
18. Betzig, E. et al. Imaging intracellular fluorescent proteins at nanometer resolution. *Science* **313**, 1642-1645 (2006).
19. Rust, M.J., Bates, M. & Zhuang, X. Sub-diffraction-limit imaging by stochastic optical reconstruction microscopy (STORM). *Nat. Meth.* **3**, 793-5 (2006).
20. Denk, W., Strickler, J.H. & Webb, W.W. Two-photon laser scanning fluorescence microscopy. *Science* **248**, 73-76 (1990).
21. Huisken, J., Swoger, J., Del Bene, F., Wittbrodt, J. & Stelzer, E.H.K. Optical sectioning deep inside live embryos by selective plane illumination microscopy. *Science* **305**, 1007-1009 (2004).
22. Möckl, L., Roy, A.R., Petrov, P.N. & Moerner, W.E. Accurate and rapid background estimation in single-molecule localization microscopy using the deep neural network BGnet. *Proc. Natl. Acad. Sci. U.S.A.* **117**, 60-67 (2020).
23. Schindelin, J. et al. Fiji: An open-source platform for biological-image analysis. *Nat. Meth.* **9**, 676-682 (2012).
24. Huang, T.S., Yang, G.J. & Tang, G.Y. A Fast Two-Dimensional Median Filtering Algorithm. *IEEE Trans. Signal Process.* **27**, 13-18 (1979).
25. Strong, D. & Chan, T. Edge-preserving and scale-dependent properties of total variation regularization. *Inverse Probl.* **19**, S165-S187 (2003).
26. Buades, A., Coll, B. & Morel, J.M. A non-local algorithm for image denoising. In *IEEE Computer Society Conference on Computer Vision and Pattern Recognition (CVPR)* 60-65 (2005).
27. Huang, X. et al. Fast, long-term, super-resolution imaging with Hessian structured illumination microscopy. *Nat. Biotechnol.* **36**, 451-459 (2018).
28. Lecun, Y., Bengio, Y. & Hinton, G. Deep learning. *Nature* **521**, 436-444 (2015).
29. Chen, J. et al. Three-dimensional residual channel attention networks denoise and sharpen fluorescence microscopy image volumes. *Nat. Methods* **18**, 678-687 (2021).

30. Fang, L. et al. Deep learning-based point-scanning super-resolution imaging. *Nat. Methods* **18**, 406-416 (2021).
31. Qiao, C. et al. Evaluation and development of deep neural networks for image super-resolution in optical microscopy. *Nat. Methods* **18**, 194-202 (2021).
32. Qiao, C. et al. Rationalized deep learning super-resolution microscopy for sustained live imaging of rapid subcellular processes. *Nat. Biotechnol.* **41**, 367-377 (2023).
33. Gustafsson, M.G.L. Surpassing the lateral resolution limit by a factor of two using structured illumination microscopy. *Journal of Microscopy* **198**, 82-87 (2000).
34. Chen, B.C. et al. Lattice light-sheet microscopy: Imaging molecules to embryos at high spatiotemporal resolution. *Science* **346**, 439-+ (2014).
35. York, A.G. et al. Instant super-resolution imaging in live cells and embryos via analog image processing. *Nat. Meth.* **10**, 1122-1130 (2013).
36. Hoebe, R.A. et al. Controlled light-exposure microscopy reduces photobleaching and phototoxicity in fluorescence live-cell imaging. *Nat. Biotechnol.* **25**, 249-253 (2007).
37. Chu, K.K., Lim, D. & Mertz, J. Enhanced weak-signal sensitivity in two-photon microscopy by adaptive illumination. *Opt. Lett.* **32**, 2846-2848 (2007).
38. Li, B., Wu, C., Wang, M., Charan, K. & Xu, C. An adaptive excitation source for high-speed multiphoton microscopy. *Nat. Methods* **17**, 163-166 (2020).
39. Staudt, T. et al. Far-field optical nanoscopy with reduced number of state transition cycles. *Opt. Express* **19**, 5644-5657 (2011).
40. Chakrova, N., Canton, A.S., Danelon, C., Stallinga, S. & Rieger, B. Adaptive illumination reduces photobleaching in structured illumination microscopy. *Biomed. Opt. Express* **7**, 4263-4274 (2016).
41. Göttfert, F. et al. Strong signal increase in STED fluorescence microscopy by imaging regions of subdiffraction extent. *Proc. Natl. Acad. Sci. U.S.A.* **114**, 2125-2130 (2017).
42. Heine, J. et al. Adaptive-illumination STED nanoscopy. *Proc. Natl. Acad. Sci. U. S. A.* **114**, 9797-9802 (2017).
43. Dreier, J. et al. Smart scanning for low-illumination and fast RESOLFT nanoscopy in vivo. *Nat. Commun.* **10**(2019).
44. Vinçon, B., Geisler, C. & Egner, A. Pixel hopping enables fast STED nanoscopy at low light dose. *Opt. Express* **28**, 4516-4528 (2020).
45. Štefko, M., Ottino, B., Douglass, K.M. & Manley, S. Autonomous illumination control for localization microscopy. *Opt. Express* **26**, 30882-30900 (2018).
46. Chiron, L. et al. CyberSco.Py an open-source software for event-based, conditional microscopy. *Sci. Rep.* **12**(2022).
47. Almada, P. et al. Automating multimodal microscopy with NanoJ-Fluidics. *Nat. Commun.* **10**(2019).
48. Edelstein, A.D. et al. Advanced methods of microscope control using µManager software. *J. Biol. Methods* **1**(2014).
49. Pinkard, H., Stuurman, N., Corbin, K., Vale, R. & Krummel, M.F. Micro-Magellan: Open-source, sample-adaptive, acquisition software for optical microscopy. *Nat. Methods* **13**, 807-809 (2016).
50. Fox, Z.R. et al. Enabling reactive microscopy with MicroMator. *Nat. Commun.* **13**(2022).
51. Casas Moreno, X. et al. An open-source microscopy framework for simultaneous control of image acquisition, reconstruction, and analysis. *HardwareX* **13**(2023).

52. Schermelleh, L. et al. Super-resolution microscopy demystified. *Nat. Cell Biol.* **21**, 72-84 (2019).
53. Wu, Y. & Shroff, H. Multiscale fluorescence imaging of living samples. *Histochem. Cell Biol.* **158**, 301-323 (2022).
54. Sarder, P. & Nehorai, A. Deconvolution methods for 3-D fluorescence microscopy images. *IEEE Signal Process. Mag.* **23**, 32-45 (2006).
55. Lucy, L.B. An iterative technique for the rectification of observed distributions. *Astron. J.* **79**, 745-754 (1974).
56. Richardson, W.H. Bayesian-based iterative method of image restoration. *J. Opt. Soc. Am.* **62**, 55-59 (1972).
57. Sage, D. et al. DeconvolutionLab2: An open-source software for deconvolution microscopy. *Methods* **115**, 28-41 (2017).
58. Zhao, W. et al. Sparse deconvolution improves the resolution of live-cell super-resolution fluorescence microscopy. *Nat. Biotechnol.* **40**, 606-617 (2022).
59. Li, Y. et al. Incorporating the image formation process into deep learning improves network performance. *Nat. Methods* **19**, 1427-1437 (2022).
60. Chhetri, R.K. et al. Whole-animal functional and developmental imaging with isotropic spatial resolution. *Nat. Methods* **12**, 1171-1178 (2015).
61. Wu, Y. et al. Spatially isotropic four-dimensional imaging with dual-view plane illumination microscopy. *Nat. Biotechnol.* **31**, 1032-1038 (2013).
62. Wu, Y. et al. Multiview confocal super-resolution microscopy. *Nature* **600**, 279-284 (2021).
63. Ingaramo, M. et al. Richardson-Lucy deconvolution as a general tool for combining images with complementary strengths. *ChemPhysChem* **15**, 794-800 (2014).
64. Kazemipour, A. et al. Kilohertz frame-rate two-photon tomography. *Nat. Methods* **16**, 778-786 (2019).
65. Wang, H. et al. Deep learning enables cross-modality super-resolution in fluorescence microscopy. *Nat. Methods* **16**, 103-110 (2019).
66. Li, X. et al. Three-dimensional structured illumination microscopy with enhanced axial resolution. *Nat. Biotechnol.* (2023).
67. Weigert, M., Royer, L., Jug, F. & Myers, G. Isotropic reconstruction of 3D fluorescence microscopy images using convolutional neural networks. in Lecture Notes in Computer Science (including subseries Lecture Notes in Artificial Intelligence and Lecture Notes in Bioinformatics) Vol. 10434 LNCS 126-134 (2017).
68. Hampson, K.M. et al. Adaptive optics for high-resolution imaging. *Nat. Rev. Methods Primers* **1**(2021).
69. Ji, N. Adaptive optical fluorescence microscopy. *Nat. Methods* **14**, 374-380 (2017).
70. Wang, K. et al. Rapid adaptive optical recovery of optimal resolution over large volumes. *Nat. Methods* **11**, 625-628 (2014).
71. Liu, T.L. et al. Observing the cell in its native state: Imaging subcellular dynamics in multicellular organisms. *Science* **360**(2018).
72. Rodríguez, C. et al. An adaptive optics module for deep tissue multiphoton imaging in vivo. *Nat. Methods* **18**, 1259-1264 (2021).
73. Streich, L. et al. High-resolution structural and functional deep brain imaging using adaptive optics three-photon microscopy. *Nat. Methods* **18**, 1253-1258 (2021).

74. Lin, R., Kipreos, E.T., Zhu, J., Khang, C.H. & Kner, P. Subcellular three-dimensional imaging deep through multicellular thick samples by structured illumination microscopy and adaptive optics. *Nat. Commun.* **12**(2021).
75. Zheng, W. et al. Adaptive optics improves multiphoton super-resolution imaging. *Nat. Methods* **14**, 869-872 (2017).
76. Saha, D. et al. Practical sensorless aberration estimation for 3D microscopy with deep learning. *Opt. Express* **28**, 29044-29053 (2020).
77. Vinogradova, K. & Myers, E.W. Estimation of Optical Aberrations in 3D Microscopic Bioimages. In *7th International Conference on Frontiers of Signal Processing, ICFSP* 97-103 (2022).
78. Feng, B.Y. et al. NeuWS: Neural wavefront shaping for guidestar-free imaging through static and dynamic scattering media. *Sci. Adv.* **9**, eadg4671 (2023).
79. Balzarotti, F. et al. Nanometer resolution imaging and tracking of fluorescent molecules with minimal photon fluxes. *Science* **355**, 606-612 (2017).
80. Deguchi, T. et al. Direct observation of motor protein stepping in living cells using MINFLUX. *Science* **379**, 1010-1015 (2023).
81. Wolff, J.O. et al. MINFLUX dissects the unimpeded walking of kinesin-1. *Science* **379**(2023).
82. Royer, L.A. et al. Adaptive light-sheet microscopy for long-term, high-resolution imaging in living organisms. *Nat. Biotechnol.* **34**, 1267-1278 (2016).
83. Pinkard, H. et al. Learned adaptive multiphoton illumination microscopy for large-scale immune response imaging. *Nat. Commun.* **12**(2021).
84. Durand, A. et al. A machine learning approach for online automated optimization of super-resolution optical microscopy. *Nat. Commun.* **9**, 5247 (2018).
85. Sheppard, C.J.R. Super-resolution in confocal imaging. *Optik (Jena)* **80**, 53-54 (1988).
86. Müller, C.B. & Enderlein, J. Image scanning microscopy. *Physical Review Letters* **104**(2010).
87. York, A.G. et al. Resolution doubling in live, multicellular organisms via multifocal structured illumination microscopy. *Nat. Methods* **9**, 749-754 (2012).
88. Schulz, O. et al. Resolution doubling in fluorescence microscopy with confocal spinning-disk image scanning microscopy. *Proc. Natl. Acad. Sci. U.S.A.* **110**, 21000-21005 (2013).
89. De Luca, G.M.R. et al. Re-scan confocal microscopy: scanning twice for better resolution. *Biomed. Opt. Express* **4**, 2644-2656 (2013).
90. Roth, S., Sheppard, C.J.R., Wicker, K. & Heintzmann, R. Optical photon reassignment microscopy (OPRA). *Opt. Nano.* **2**, 1-6 (2013).
91. Azuma, T. & Kei, T. Super-resolution spinning-disk confocal microscopy using optical photon reassignment. *Opt. Express* **23**, 15003-15011 (2015).
92. Hofmann, M., Eggeling, C., Jakobs, S. & Hell, S.W. Breaking the diffraction barrier in fluorescence microscopy at low light intensities by using reversibly photoswitchable proteins. *Proc. Natl. Acad. Sci. U. S. A.* **102**, 17565-17569 (2005).
93. Chmyrov, A. et al. Nanoscopy with more than 100,000 'doughnuts'. *Nat. Methods* **10**, 737-740 (2013).
94. Bodén, A. et al. Volumetric live cell imaging with three-dimensional parallelized RESOLFT microscopy. *Nat. Biotechnol.* **39**, 609-618 (2021).
95. Lelek, M. et al. Single-molecule localization microscopy. *Nat. Rev. Methods Primers* **1**(2021).

96. Speiser, A. et al. Deep learning enables fast and dense single-molecule localization with high accuracy. *Nat. Methods* **18**, 1082-1090 (2021).
97. Mahecic, D. et al. Event-driven acquisition for content-enriched microscopy. *Nat. Methods* **19**, 1262-1267 (2022).
98. Alvelid, J., Damenti, M., Sgattoni, C. & Testa, I. Event-triggered STED imaging. *Nat. Methods* **19**, 1268-1275 (2022).
99. Tsien, R.Y. The green fluorescent protein. *Annu. Rev. Biochem.* **67**, 509-544 (1998).
100. Lambert, T.J. FPbase: a community-editable fluorescent protein database. *Nat. Methods* **16**, 277-278 (2019).
101. Cheng, S. et al. Single-cell cytometry via multiplexed fluorescence prediction by label-free reflectance microscopy. *Sci. Adv.* **7**(2021).
102. Christiansen, E.M. et al. In Silico Labeling: Predicting Fluorescent Labels in Unlabeled Images. *Cell* **173**, 792-803.e19 (2018).
103. McRae, T.D., Oleksyn, D., Miller, J. & Gao, Y.-R. Robust blind spectral unmixing for fluorescence microscopy using unsupervised learning. *PLOS ONE* **14**, e0225410 (2019).
104. Seo, J. et al. PICASSO allows ultra-multiplexed fluorescence imaging of spatially overlapping proteins without reference spectra measurements. *Nat. Commun.* **13**, 2475 (2022).
105. Digman, M.A., Caiolfa, V.R., Zamai, M. & Gratton, E. The Phasor Approach to Fluorescence Lifetime Imaging Analysis. *Biophys. J.* **94**, L14-L16 (2008).
106. Lanzanò, L. et al. Encoding and decoding spatio-temporal information for super-resolution microscopy. *Nat. Commun.* **6**, 6701 (2015).
107. Scipioni, L., Rossetta, A., Tedeschi, G. & Gratton, E. Phasor S-FLIM: a new paradigm for fast and robust spectral fluorescence lifetime imaging. *Nat. Methods* **18**, 542-550 (2021).
108. Smith, J.T., Ochoa, M. & Intes, X. UNMIX-ME: spectral and lifetime fluorescence unmixing via deep learning. *Biomed. Opt. Express* **11**, 3857-3874 (2020).
109. Belthangady, C. & Royer, L.A. Applications, promises, and pitfalls of deep learning for fluorescence image reconstruction. *Nat. Methods* **16**, 1215-1225 (2019).
110. Saguy, A. et al. DBlink: dynamic localization microscopy in super spatiotemporal resolution via deep learning. *Nat. Methods* (2023).
111. Milias-Argeitis, A. et al. In silico feedback for in vivo regulation of a gene expression circuit. *Nat. Biotechnol.* **29**, 1114-1116 (2011).
112. Emiliani, V., Cohen, A.E., Deisseroth, K. & Häusser, M. All-optical interrogation of neural circuits. *J. Neurosci.* **35**, 13917-13926 (2015).
113. Schmidt, M. & Lipson, H. Distilling Free-Form Natural Laws from Experimental Data. *Science* **324**, 81-85 (2009).
114. Colón-Ramos, D.A., La Riviere, P., Shroff, H. & Oldenbourg, R. Transforming the development and dissemination of cutting-edge microscopy and computation. *Nat. Methods* **16**, 667-669 (2019).
115. Conrad, C. et al. Micropilot: Automation of fluorescence microscopy-based imaging for systems biology. *Nat. Methods* **8**, 246-249 (2011).
116. Wei, O. et al. BioImage Model Zoo: A Community-Driven Resource for Accessible Deep Learning in BioImage Analysis. *bioRxiv*, 2022.06.07.495102 (2022).
117. Katona, G. et al. Fast two-photon in vivo imaging with three-dimensional random-access scanning in large tissue volumes. *Nat. Methods* **9**, 201-208 (2012).

118. Krull, A., Buchholz, T.O. & Jug, F. Noise2void-Learning denoising from single noisy images. In *Proceedings of the IEEE Computer Society Conference on Computer Vision and Pattern Recognition* 2124-2132 (2019).
119. Ouyang, W., Aristov, A., Lelek, M., Hao, X. & Zimmer, C. Deep learning massively accelerates super-resolution localization microscopy. *Nat. Biotechnol.* **36**, 460-468 (2018).
120. Greener, J.G., Kandathil, S.M., Moffat, L. & Jones, D.T. A guide to machine learning for biologists. *Nat. Rev. Mol. Cell Biol.* **23**, 40-55 (2022).
121. Wang, Z., Bovik, A.C., Sheikh, H.R. & Simoncelli, E.P. Image quality assessment: From error visibility to structural similarity. *IEEE Transactions on Image Processing* **13**, 600-612 (2004).
122. Wang, Z., Simoncelli, E.P. & Bovik, A.C. Multiscale structural similarity for image quality assessment. In *The Thrity-Seventh Asilomar Conference on Signals, Systems & Computers, 2003* 1398-1402 Vol.2 (2003).
123. Zhang, R., Isola, P., Efros, A.A., Shechtman, E. & Wang, O. The Unreasonable Effectiveness of Deep Features as a Perceptual Metric. In *2018 IEEE/CVF Conference on Computer Vision and Pattern Recognition* 586-595 (2018).
124. Lehtinen, J. et al. Noise2Noise: Learning image restoration without clean data. In *35th International Conference on Machine Learning, ICML 2018* 4620-4631 (2018).
125. Kefer, P. et al. Performance of deep learning restoration methods for the extraction of particle dynamics in noisy microscopy image sequences. *Molecular Biology of the Cell* **32**, 903-914 (2021).
126. Wu, Y. et al. Reflective imaging improves spatiotemporal resolution and collection efficiency in light sheet microscopy. *Nat. Commun.* **8**(2017).
127. Shaevitz, J.W. & Fletcher, D.A. Enhanced three-dimensional deconvolution microscopy using a measured depth-varying point-spread function. *J. Opt. Soc. Am. A* **24**, 2622-2627 (2007).
128. Yanny, K., Monakhova, K., Shuai, R.W. & Waller, L. Deep learning for fast spatially varying deconvolution. *Optica* **9**, 96-99 (2022).
129. Guo, M. et al. Rapid image deconvolution and multiview fusion for optical microscopy. *Nat. Biotechnol.* **38**, 1337-1346 (2020).

Acknowledgements

We thank Grant Kroeschell and Richard Ikegami (Shroff lab) and Jiji Chen (NIH Advanced Imaging and Microscopy Resource) for help with figure preparation. This work was supported by the Howard Hughes Medical Institute (HHMI) and the École Polytechnique Fédérale de Lausanne (EPFL).

Author contributions

The authors contributed equally to all aspects of the article.

Competing interests

H.S. is co-inventor on US patent 9,696,534, owned by NIH and licensed to VisiTech International and Yokogawa Electric Corporation, describing multifocal and analog implementations of SIM, including the instant SIM mentioned here. H.S. has also filed invention disclosures on 4-beam SIM and multiview confocal microscopy, which both rely on the deep learning strategies mentioned here.

**Glossary, terms are defined in the context of fluorescence microscopy**

| | |
|---|---|
| aberrations | Distortions in the images generated by an optical system due to deviations in the properties of real optical components such as lenses, mirrors, and filters, as compared with theoretical models; or due to refractive index variations in the sample. |
| artificial neural network (ANN) | A computational model composed of interconnected nodes and layers, designed to mimic the structure and function of the brain. |
| bleed through | Overlap in the emission spectra of two distinct fluorophores, leading to detection of both at the same wavelength. |
| binning | The denoising process of combining data from adjacent pixels in an image, which results in fewer pixels. |
| centroid | The center position of an object, corresponding to the weighted mean of pixel intensity values. |
| continuity (spatial, temporal) | Property of an object of containing no resolved gaps in space (spatial continuity) or time (temporal continuity). |
| crosstalk | Undesired mixing of signals. For example, overlap in the excitation spectra of two distinct fluorophores, leading to excitation of both by the same wavelength of light. Or, in multifocal microscopy, fluorescence from one excitation spot contributing to the signal of neighboring regions. |
| deep learning | A class of machine learning algorithms based on artificial neural networks (ANN) containing multiple data processing layers. |
| depletion donut | A donut-shaped illumination light which is used to turn off the fluorescence in the periphery of the focal spot. |
| denoising | Image processing procedure to remove noise. |
| downsampling | Reducing the sampling rate, e.g. spatially or temporally. |
| dwell time | The time a focused laser beam is applied to each location in the specimen being imaged. |
| fluorescence lifetime | The time a fluorophore spends in the excited state before emitting a photon and returning to the ground state. |
| Gaussian readout noise | Noise which follows a Gaussian distribution and is independent of pixel intensity values, e.g. noise generated by a camera chip when it converts charge into voltage. |
| ground truth | The target of a deep learning model, for example, a label against which the predictions of a model are compared during training. |
| hallucination | Neural network output which looks plausible, but has no basis in the input data. |
| light sheet microscopy | Method in which a thin slice of the specimen is illuminated perpendicular to the imaging orientation. |
| lattice light sheet (LLS) | Light sheet generated by scanning a two-dimensional lattice of structured light known as Bessel beams. |
| linearity | The property of two quantities (e.g., intensities) being linearly proportional, such that their values are related by a multiplicative constant. |
| low-pass filter | An operation that passes frequencies below a cutoff value, which corresponds to retaining lower resolution features in an image. |
| mean squared error (MSE) | The mean squared difference between measured and predicted values, e.g. ground truth pixel intensity values and those output by a network. Used as a metric of how well a model captures the data. |
| median filter | An operation that replaces the intensity value in a pixel by the median value of its neighbors. |
| multi-photon microscopy | Method in which multiple photons must be simultaneously absorbed by a single fluorophore to bring it into its excited state. |
| Nyquist-Shannon sampling theorem | Principle that defines the maximum spacing between measurements that will be sufficient to determine an underlying signal of a given frequency. E.g., to resolve dynamics at a timescale of T seconds, the time between images should be less than T/2. |
| peak signal to noise ratio (PSNR) | The ratio between the squared maximum possible signal in an image and mean squared error, defined above. This is reported in units of decibels, so the logarithm of the ratio is taken and multiplied by 10. Used as a metric of how well a model captures the data. |
| photon budget | Number of photons detected from an object of interest during an experiment. |
| phototoxicity | Damage to living specimens caused by exposure to light. |
| point spread function | The intensity distribution of a point-like source when imaged through a microscope. |
| Poisson noise | Noise (i.e., shot noise) which follows a Poisson distribution, e.g. arising from measuring photons because they are discrete particles. |
| pyramid of frustration | Schematic illustrating the tradeoffs in fluorescence microscopy, where each axis defines one measurement property such as SNR, spatial, or temporal resolution. The fixed photon budget implies that improving along one dimension degrades along another. |
| reference dataset | Data used within a field to compare the performance of algorithms, in benchmarking comparisons. |
| reversible saturable optical fluorescence transitions (RESOLFT) | A super resolution technique suitable for live cell imaging and based on reversibly switching fluorescent probes and patterned illumination |
| sampling | Recording a signal in a discontinuous manner, at specific locations or times. |
| synthetic data (and semi-synthetic) | Data generated computationally using a model. Can be combined with real data to create semi-synthetic data. |
| signal to noise ratio (SNR) | The ratio between signal and noise, which can be estimated on a per-pixel basis as the mean intensity value divided by the standard deviation of the intensity. |
| single molecule localization microscopy (SMLM) | A class of super-resolution microscopy techniques based on imaging single molecules whose signals have been isolated, then combining their sub-pixel locations to form a composite image. |
| spatial frequency | Just as a function can be decomposed into a sum of sines and cosines (cf. Fourier transform), an image can be decomposed into a sum of waves with different spatial frequencies. These represent the image at different resolution levels, with higher spatial frequencies describing finer image details and lower spatial frequencies describing coarser details. |
| spatial resolution | The smallest distance at which two features can be distinguished. |
| spectrum (emission, excitation) | Response of a fluorophore as a function of wavelength of light. The excitation spectrum reflects the capacity of light absorbed at different wavelengths to generate fluorescence of a particular wavelength, while the emission spectrum reflects the emission of light at different wavelengths following excitation at a particular wavelength. |
| spinning disk confocal microscopy | Method in which an array of focused excitation laser beams is produced by an array of microlenses on a disk which spins to scan the specimen. Out-of-focus emission light is rejected by a confocal array of pinholes. |
| statistical distance | An objective score that summarizes statistical differences between two objects, for example, between a prediction and the training set in machine learning. Possibilities include “total variational distance” and “Kullback-Liebler divergence.” |
| stimulated emission depletion (STED) microscopy | A super resolution technique based on patterned illumination, typically donut shaped, which is used to deplete the fluorescence of commonly used fluorescent probes |
| structural similarity index measure (SSIM) | A measure of how similar two images are, based on distortions which humans tend to perceive: the weighted product of luminance (average brightness), contrast (standard |
| | deviation of pixel intensity values), and structure (cross-covariance). |
| structured illumination microscopy (SIM) | A class of super-resolution microscopy techniques which use patterned excitation light combined with optical or digital image processing to recover information below the diffraction limit. |
| super-resolution | Imaging techniques which achieve spatial resolutions surpassing the diffraction limit of light. |
| supervised learning | Machine learning training process which takes user-assigned labels as the ground truth, to compare with algorithm predictions. |
| temporal resolution | The time between consecutive images of the same part of the specimen. |
| total internal reflection fluorescence (TIRF) | Method in which a specimen is illuminated at the coverslip-media interface by an evanescent field generated by a laser at an incident angle sufficient to cause total internal reflection. |
| training data | Data used to train a machine learning algorithm to make predictions. |

## Table 1: Summary of methods

| Method | Type: classical [CL], deep learning [DL], hardware [H] | Advantages | Requirements / assumptions | Drawbacks, potential artifacts | Proof of concept method or system | Availability |
|---|---|---|---|---|---|---|
| **Denoising** | | | | | | |
| binning | CL | simple | object is oversampled by the detector | reduction in the spatial resolution | general | common (ImageJ, FIJI, etc.) |
| median filtering [24] | CL | simple | intensity of each pixel is related to that of its neighbors | smooths signal, removes edges | general | common (ImageJ, FIJI, etc.) |
| total variation regularization [25] | CL | preserves edges | noise corresponds to excessive variation in the image | creation of patches of pixels with the same intensity, “staircase artifacts” | general | e.g., existing SciPy function (denoise_tv_chambolle) |
| non-local means [26] | CL | considers values of other pixels in the image | user-defined weighting function that relates different pixels | computationally expensive | general | FIJI (https://imagej.net/plugins/non-local-means-denoise/), napari (https://www.napari-hub.org/plugins/napari-nlm), Skimage (https://scikit-image.org/docs/stable/auto_examples/filters/plot_nonlocal_means.html), etc. |
| Hessian SIM [27] | CL | effective at very low SNR | assumes spatial and temporal continuity of objects | computationally expensive | 2D SIM, organelles, cytoskeleton | GitHub pycudasirecon (https://github.com/tlambert03/pycudasirecon) |
| content-aware restoration (CARE) [11] | DL (supervised) | With proper training, one can expect resolution improvements ~2-3 fold over input; can denoise so that illumination intensity can be lowered more than ~10 fold over conventional imaging. | requires training data | Fails especially at ultra low input SNR (~< 3); or > 2-fold resolution enhancement. Depending on the method results are either blurry, or start hallucinating structures (e.g., in background regions)<br><br>Slow to train (hours, longer than image acquisition)<br><br>Quality of results depends on similarity between data and training data. | small organisms, tissues, cytoskeleton | CSBDeep (https://csbdeep.bioimagecomputing.com)<br>GitHub CSBDeep (https://github.com/CSBDeep/CSBDeep) |
| residual channel attention network (RCAN) restoration [29] | | | | | 3D organelles, cytoskeleton, chromosomes | GitHub 3D-RCAN (https://github.com/AiviaCommunity/3D-RCAN) |
| deep Fourier channel attention network (DFCAN) restoration [31] | | | | | 2D SIM, cytoskeleton, organelles | GitHub DL-SR (https://github.com/qc17-THU/DL-SR) |
| point scanning super-resolution (PSSR) [30] | | effective for under-sampled, noisy data | point scanning data, training data | | confocal, EM, organelles | 3DEM.org |
| rationalized deep learning (rDL) [32] | | improves prediction uncertainty | prior knowledge of illumination patterns, training data | | 2D SIM, 3D LLS, cytoskeleton, organelles, chromosomes | GitHub rDL-SIM (https://github.com/qc17-THU/rDL-SIM) |
| Noise2Void | DL (unsupervised) | removes pixel noises (e.g. Poisson and gaussian readout noise), no training or low-noise data required | noise must be pixel-independent | all spatially correlated signals are retained, including imaging artifacts | general | FIJI (https://imagej.net/plugins/n2v)<br>napari (https://www.piwheels.org/project/napari-n2v/)<br>GitHub n2v (https://github.com/juglab/n2v) |
| controlled light exposure microscopy (CLEM) [36] | CL + H | improves SNR at reduced excitation light dose | high signal regions are the most important | relative intensities in different areas are difficult to interpret | confocal, cytoskeleton, organelles | no |
| **spatial resolution** | | | | | | |
| Richardson-Lucy deconvolution | CL | improves contrast and SNR as well | PSF is a user input, assumed constant across image | too many iterations lead to over-sharpening, light and dark rings may appear around objects | general | common (ImageJ, FIJI, etc) |
| Sparse-SIM [58] | CL | increases resolution at high frame rate | sparsity and continuity of structures | distortions and intensity fluctuations at low SNR | SIM, spinning disk confocal, cytoskeleton, organelles | GitHub Sparse-SIM (https://github.com/WeisongZhao/Sparse-SIM) |
| Richardson-Lucy Network (RLN) [59] | DL (supervised) | improves performance axially and at low SNR, faster computation | assumes that RL is a good model for deconvolution, training data | distortions at lowest SNR | tissues, embryos, cells | GitHub Richardson-Lucy-Net (https://github.com/MeatyPlus/Richardson-Lucy-Net) |
| Cross-modality super-resolution [65] | DL (supervised) | high-resolution modality from low-resolution | information within low-resolution images is sufficient, training data | in the absence of sufficient information in low-resolution images, network may generate hallucinations | confocal, STED, cytoskeleton, organelles | FIJI/ImageJ plugin |
| 3D SIM restoration [66] | DL (supervised) | restores axial resolution based on higher resolution lateral views | objects should appear similar viewed laterally or axially, training data | time-consuming to collect volumetric data and train, prone to distortions at low SNR | 3D SIM, bacteria, cytoskeleton, organelles | GitHub ( https://github.com/eexuesong/SIMreconProject) |
| Multiphoton AO [72] | CL + H | corrects aberrations for deep imaging | indirect wavefront sensing of aberrations | setup includes expensive lasers, sensing slower than image acquisition | multiphoton, mouse brain | GitHub JiLabAO (https://github.com/JiLabUCBerkeley/JiLabAO, https://github.com/prevedel-lab/AO.git) |
| SIM AO [74] | CL + H | corrects aberrations for deeper super-resolution | indirect wavefront sensing of aberrations | high level of user expertise | SIM, C elegans, plants, cells | GitHub (https://github.com/Knerlab) |
| Multiphoton SIM AO [75] | CL + H | corrects aberrations for deeper super-resolution | direct wavefront sensing of aberrations | setup includes expensive lasers, sensing ~1 s or longer, point-scanning acquisition is slow | multiphoton iSIM, embryos, cells | |
| PHASENET [76] | DL (supervised) | corrects aberrations for deep imaging | aberration estimation from synthetic data, training data | only demonstrated on beads | confocal, widefield, beads | GitHub (https://github.com/mpicbg-csbd/phasenet) |
| MINFLUX [79] | CL + H | requires fewer photons to localize molecules | single molecules are isolated | high level of user expertise, complex optical setup | specialized microscope, small molecular structures | |
| AutoPilot [82] | CL + H | matches excitation and imaging planes in real time | sample changes between timepoints are small | high level of user expertise, complex optical setup | specialized microscope, embryos | GitHub (https://microscopeautopilot.github.io/) |
| Learned adaptive multiphoton illumination (LAMI) [83] | DL + H (supervised) | optimizes local excitation in 3D sample | requires specific calibration dataset for training, training data | high level of user expertise, strong reliance on model | multiphoton, lymph node | Zenodo (https://doi.org/10.5281/zenodo.4314107) |
| **temporal resolution** | | | | | | |
| Multifocal SIM [87] | CL + H | several cells thick, 1 Hz imaging speed | access to custom or commercial microscope | complex optical setup | specialized microscope, embryos, cells, cytoskeleton | Google Code (http://code.google.com/p/msim/) |
| Spinning disk confocal ISM [88] | CL + H | implemented on commercial microscope, | advanced programming capability | customized control software | spinning disk, cytoskeleton, molecular complexes | |
| | | 1-3 Hz imaging speed | | | | |
| instant SIM (iSIM) [35] | CL + H | real-time optical image processing, 100 Hz imaging speed | access to custom or commercial microscope | complex optical setup | specialized microscope, embryos, cells, cytoskeleton, organelles | Google Code (http://code.google.com/p/msim/) |
| Parallelized RESOLFT [93] | CL + H | higher resolution, 1 Hz imaging speed | reversibly switchable fluorophores | complex optical setup, sensitive to crosstalk | specialized microscope, cytoskeleton | Image reconstruction software included with manuscript |
| 3D pRESOLFT [94] | CL + H | higher resolution, 1-2 Hz imaging speed, 3D | reversibly switchable fluorophores | complex optical setup | specialized microscope, cytoskeleton, synaptic proteins, organelles | Acquisition software available at Github (https://github.com/TestaLab) |
| Multiphoton 3D random access scanning [117] | CL + H | thick samples (~1 mm), 100 Hz time resolution | sparse regions of interest | complex optical setup and control software | specialized microscope, synaptic activity in brain slices | included with manuscript |
| Event-driven acquisition (EDA) [97] | DL + H | on-demand shift in frame rates | events of interest are rare and have precursors, training data | measurement bias toward events detected by network, complex setup | iSIM, widefield, organelles, bacteria | Micro-manager plugin (https://pypi.org/project/eda-plugin/) |
| Event-triggered STED (etSTED) | DL + H | on-demand shift from widefield to STED | events of interest are rare and have precursors, training data | measurement bias toward events detected by network, complex setup | STED, organelles | GitHub (https://github.com/jonatanalvelid/etSTED-widget-base) |
| **Multicolor imaging** | | | | | | |
| Learning Unsupervised Means of Spectra (LUMoS) [103] | DL (unsupervised) | no pre-knowledge needed of fluorophores | each pixel contains only one fluorophore, emission spectra share same shape | up to six fluorophores | multiphoton, cytoskeleton, organelles | figshare (https://doi.org/10.6084/m9.figshare.c.4537607) |
| Phasor spectral fluorescence lifetime imaging microscopy (FLIM) [107] | CL | avoids fitting to retrieve fluorescence lifetime, no pre-knowledge needed of fluorophores | fast electronic card for photon counting | complex hardware, customized analysis pipelines | fluorescence lifetime imaging microscopy, organelles | included with manuscript |
| UNMIX-ME (multiple emissions) | DL (supervised) | combines spectral and lifetime signatures | synthetic training data, knowledge of spectra required | requires a combination of Matlab and Python (Tensorflow) | fluorescence lifetime imaging microscopy, organs | GitHub (https://github.com/jasontsmith2718/UNMIX-ME) |

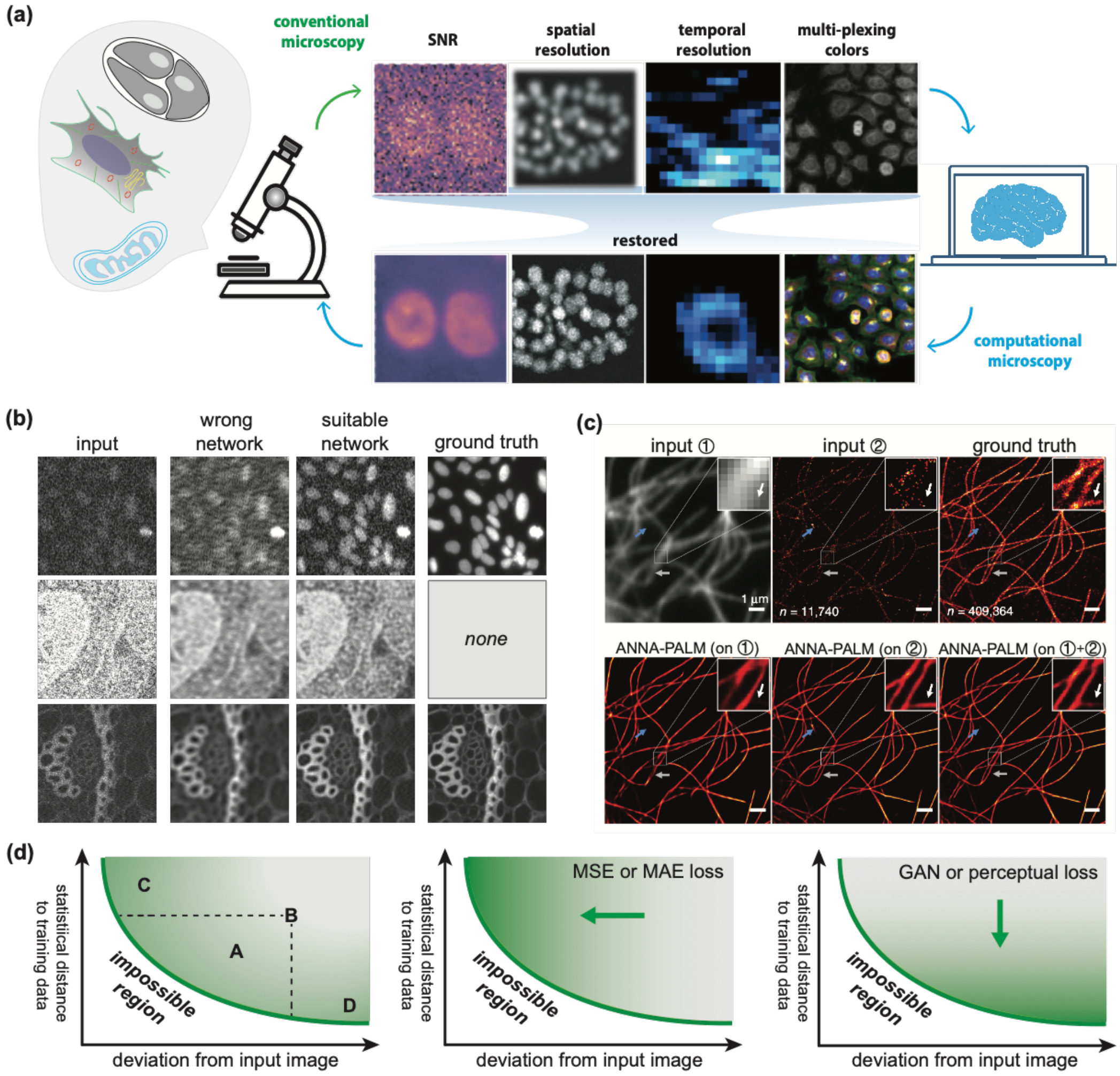

**Figure 1** | Live Cell Imaging and Deep learning: overview, examples, and limitations. **(a)** Schema of conventional and computational microscopy. A limited photon budget defines the properties of live-cell microscopy images; computational microscopy can partially circumvent the limitations, or use a feedback loop to acquire better data. **(b)** Denoising. A Noise2Void [118] network was trained on data dissimilar to the input data (wrong network) or on data similar to the input data (suitable network), for three different sample types. High-SNR ground truth is not required to train these unsupervised models and for the data in the second row does not even exist. Training the 'wrong model' with an MSE loss gives it a mismatched structural prior and results in a blurry predicted image. **(c)** Deep learning for super-resolution. Top row: widefield (input 1) and two PALM images, created by collecting a few (input 2), or many more (ground truth) single molecule events. Bottom row: ANNA-PALM predictions obtained by providing the network with either one or both input images. Figure adapted from [119]. **(d)** Image restoration takes an input image and makes it more

similar to the training data or to the ground truth image, or both (shifts it towards greener area). Since noisy or distorted images have lost information compared to ground truth, they cannot be fully restored, giving rise to an “impossible region” beyond the green line. How do the four models A, B, C, and D (leftmost plot) compare against each other? Model A is better than model B in both dimensions, but which of the models A, C, and D is “better” depends on a user’s needs. Predictions will tend to be closer to the input data (pushed towards y-axis) when using MSE or MAE losses as in CARE or Noise2Void (middle plot). Or they will tend to be closer to structures contained in the training data-when perceptual or GAN losses are used (rightmost plot). The green arrows and gradients show the direction towards different losses optimized during network training.

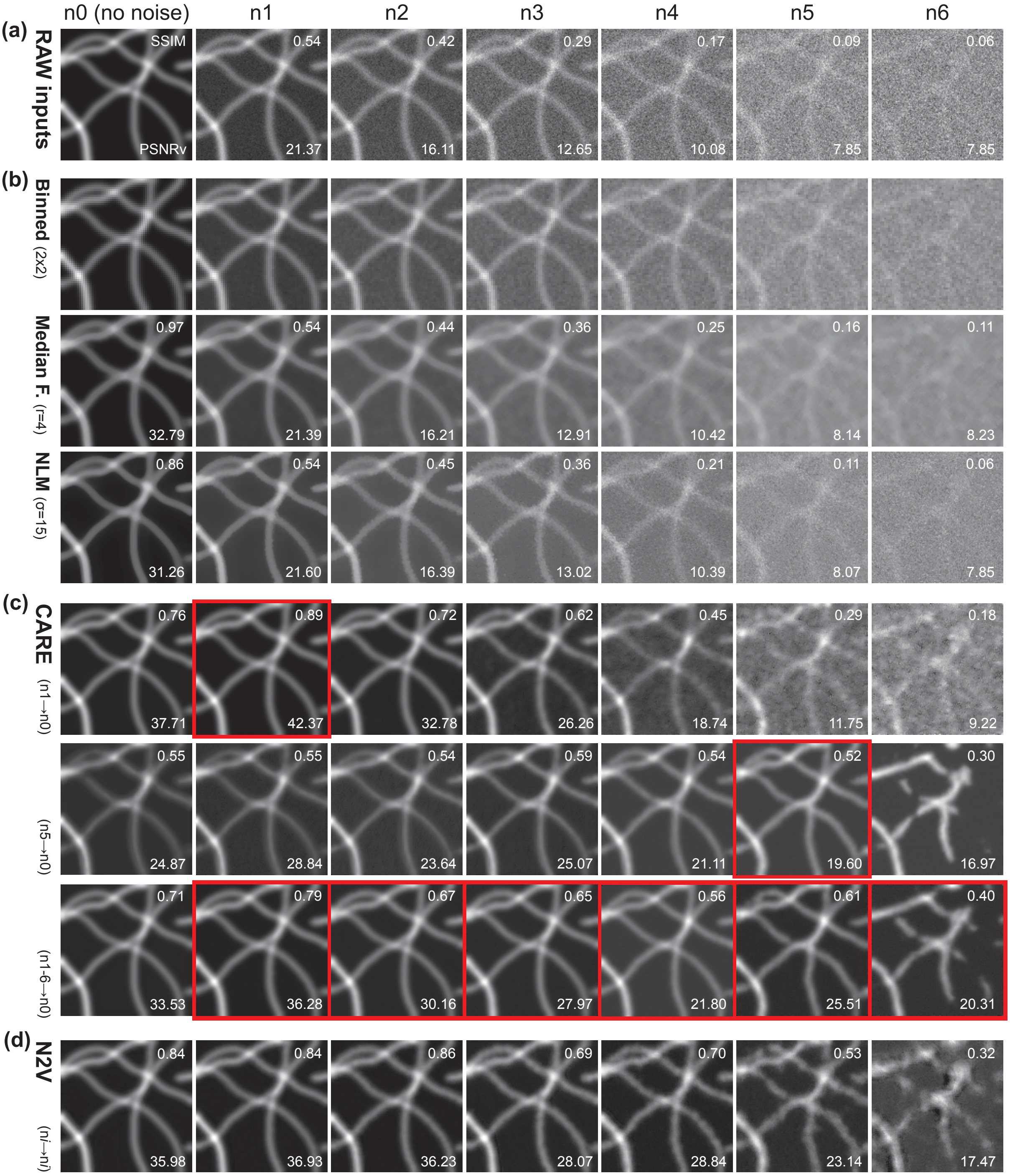

**Figure 2 | Comparison of Denoising Methods. (a)** Synthetic raw input image without noise (n0) and with varying levels of added noise (n1-n6). The noise free image is also the reference image to compute SSIM and PSNR values for all applicable images in this figure, see **Box 2** for details. **(b)** Denoising results obtained via 2x2 pixel binning (top row), median filtering with a radius of 4 pixels (middle row), and non-local means using a sigma of 15 (bottom row). **(c)** Denoising results obtained with the supervised deep learning method CARE. The three rows differ in the way the CARE model that predicted them was trained. All models used the same noise free images as ground truth but used different noise levels as inputs (red boxes). Note that the last row was trained on all pairs of non-zero noise levels n1 to n6 and then applied to all noise levels including n0. **(d)** Denoising results obtained with the self-supervised method Noise2Void. Each column shows the result obtained with a Noise2Void network

trained on this noise level specifically (no ground truth is required, except for computing the shown SSIM and PSNR values). Supervised training typically leads to better results whenever the suitable training data is available and used. Since unsupervised methods can be trained on the same body of data to be denoised, the problem of suitable training data does ~~virtually~~ not exist.

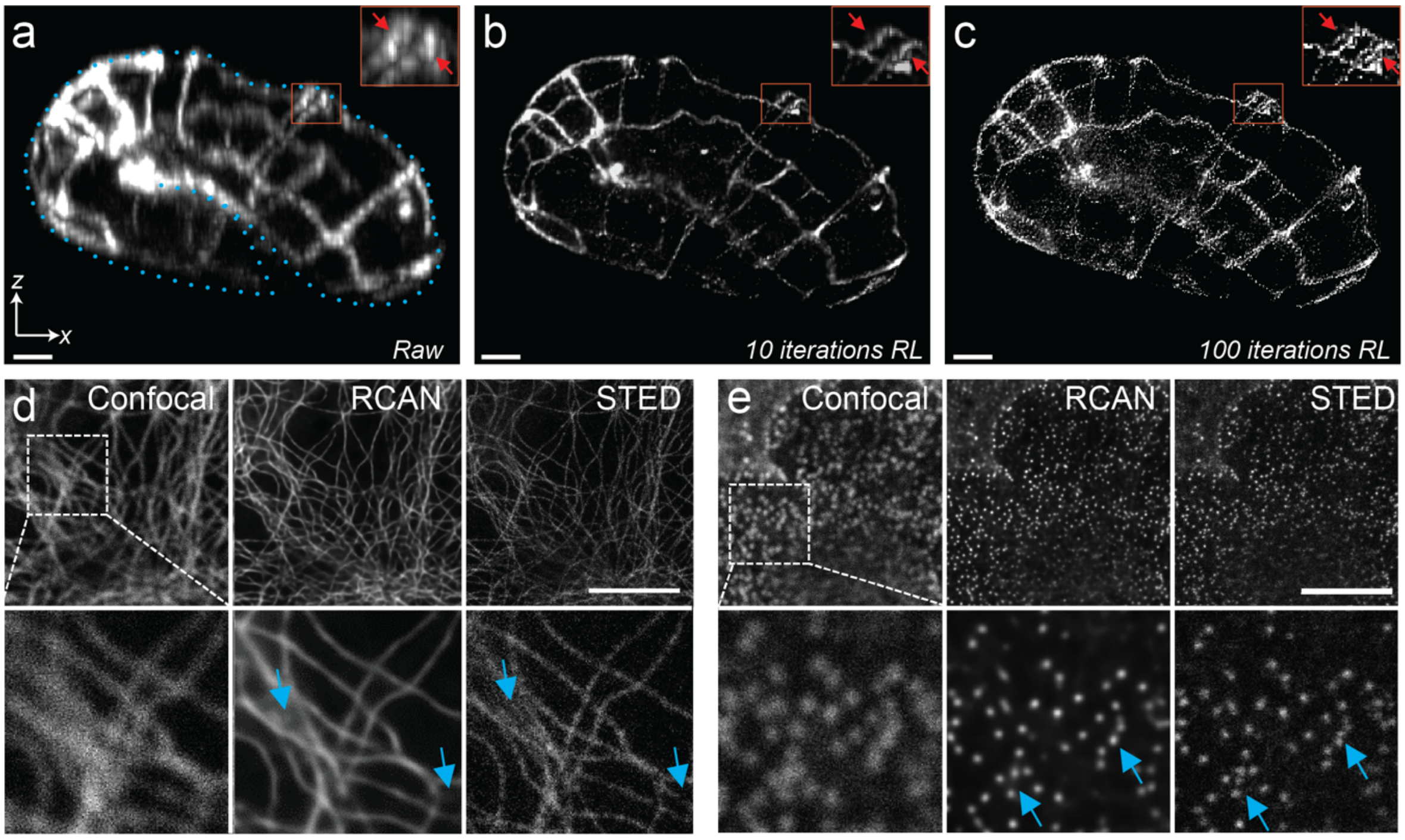

**Figure 3 | Deconvolution and deep learning for image restoration. a)** *C. elegans* embryos expressing an mScarlet junctional marker labeling the interfaces between hypodermal cells were imaged with dual-view light-sheet microscopy. Axial view maximum intensity projection of raw single-view data is shown, embryo boundary outlined with dotted blue lines. Inset shows higher magnification view of junctional markers in orange rectangular regions, with red arrows highlighting neighboring interfaces. **b)** As in **a),** but after 10 iterations of Richardson-Lucy deconvolution (RL). Note clear improvement in resolution. **c)** As in **a)**, but after 100 iterations of RL. Note amplification of noise and resulting degradation in image quality, and spurious structures created by 'oversharpening' interfaces. **d)** Immunolabeled microtubules in fixed mouse embryonic fibroblasts were imaged with confocal microscopy (left), STED microscopy (right), or restored with a residual channel attention network (RCAN) trained to predict STED images from confocal input (middle). Note obvious improvement in resolution with RCAN prediction or STED ground truth, but closer inspection (bottom row) reveals that some fine details are lost in the RCAN prediction (blue arrows). **e)** As in **d)**, but now examining immunolabeled nuclear pore complexes. While RCAN and STED ground truth both reveal more pores than confocal microscopy, RCAN does not accurately predict all pores (blue arrows, bottom row). All scale bars: 5 μm. Panels **d, e)** are modified from [29].

**Box 1: Deep Learning**

*Deep Learning* is a popular field in *Machine Learning* that uses *Artificial Neural Networks* (ANNs) to learn complex functions from data. When suitable data is available, large (deep) ANNs can learn to become powerful predictors, capable of outperforming all other known methods in a plethora of tasks, e.g. image classification, denoising, super-resolution, or segmentation.

An ANN is a sequence of *layers* containing *nodes* analogous to neurons, and *connections* between nodes in neighboring layers. Each connection between two nodes has a *weight*, roughly modeling a synapse between two neurons. During training of an ANN we are typically only changing those weights and are not adding or removing connections.

We distinguish *supervised* from *self-supervised* training depending on what data is available/required to train an ANN. Data for supervised training comes in pairs of inputs and corresponding desired output or prediction (*ground truth*). The availability of ground truth (GT) allows us to compute the error, referred to as the *loss*, between the network's current prediction and the desired result (GT). A technique known as *error backpropagation* enables us to change all weights in the network in such a way that the same input will lead to a prediction that is closer to the GT we computed the loss against. Hence, if training is successful, the loss will diminish over time and the ANN's predictions will become better. Training data must be collected in addition to data for validation of the ANN or for interpretation in a study. This is done to make sure that the network has generalized to the data type, and to avoid "overfitting," in which weights are tuned to features specific to the training set. The size of a training dataset varies, and may lie between tens and hundreds of image volumes or images.

Self-supervised training, in contrast, does not need any ground truth data, making such approaches very user-friendly. Instead, self-supervised methods change or hide some aspects of the available data and ask the ANN to predict those aspects. For example: masking single pixels and then asking a network to predict the intensities of the masked pixels trains rather powerful denoising networks. No GT is required, the original image data is enough to self-supervise the network on the desired task. While self-supervised approaches are highly sought after, for tasks other than denoising and segmentation practical unsupervised approaches remain to be developed.

If interested readers would like to dive deeper into the details of deep learning in the context of tasks relevant for biologists, we highly recommend "*A guide to machine learning for biologists*" [120].

**Box 2: Quality Measures**

In microscopy image analysis, accurately assessing the improvements made to images by classical methods or Artificial Neural Networks (ANN) is crucial. This box offers an overview of the most commonly used quality metrics, providing insights into what they measure and their appropriate applications.

Common quality measures are PSNR, SSIM, MSE, and MAE. All four require clean (ground truth quality) images, corresponding lower-quality images, and their improved, e.g. denoised, counterparts. The comparison of a restored image to the ground truth is meaningful only relative to the comparison of the corresponding unrestored image to the ground truth. For self-supervised methods, such as Noise2Void [118], such ground truth data might not exist, rendering all four metrics inapplicable.

**Mean Squared Error (MSE)** and **Mean Absolute Error (MAE)** are simple, pixel-wise error metrics. MSE calculates the squared difference, and MAE the absolute difference, between the evaluated image (*x*) and the ground truth image (*y*). Lower values indicate better image quality.

**Peak Signal-to-Noise Ratio (PSNR)**, expresses the ratio between the brightest pixel in *y* (or the brightest minus the darkest pixel in *y*) squared, and the square root of the MSE between images *x* and *y*. The final PSNR value is then ten times the logarithm (base 10) of this number, which assures that PSNR is positive, with higher values signifying a higher quality image *x*.

In many publications, the PSNR implementation additionally allows for *x* to be shifted in intensity to minimize the MSE between it and the ground truth image y. This effectively adds a constant intensity to all pixels in *x* such that the computed PSNR is maximized and the structure in the image is evaluated rather than the absolute intensity values.

The **Structure Similarity Index Measure (SSIM)** was initially developed to measures the perceptual difference between two similar looking images [121]. Explaining the details of this measure goes well beyond the scope of this box. Higher SSIM values signify better quality input images, with 1 being a perfect match between *x* and *y* (and -1 signifying perfectly anti-correlated images).

A more advanced form of SSIM, called Multiscale SSIM (MS-SSIM) attempts to measure SSIM over multiple spatial scales through a process of multiple stages of sub-sampling [122] and is sometimes used in the context of microscopy data denoising or super-resolution.

**Perceptual image quality metrics** have recently become popular in the domain of natural images [123], but we do not recommend using such measures in the context of scientific image data where the perceptual similarity between images is less important than quantifiable similarity. In simple terms, a reconstructed or restored image that is convincingly similar to real data is not the goal of scientific image processing – rather, the goal, is to arrive at an image that is as close as possible to the ground truth of the particular image.

**Box 3: Practical considerations for deep learning based denoising**

When using neural networks to denoise images, the first decision is whether to use a "supervised" or "unsupervised" approach. Supervised methods require high-quality training data, i.e., the collection of matched low SNR and high SNR image pairs. Ideally, these pairs should be perfectly registered in space and time; thus, training data is often acquired by imaging fixed samples. Although often effective, this method assumes a close correspondence between fixed and live samples, which may not be true in practice. Alternatively, training data may be gathered in live samples, acquiring the paired low- and high-SNR training pairs in rapid succession, to avoid motion-induced artifacts in the training data. For 3D applications, one useful approach is to image each focal plane in low- and high-light conditions before progressing to the next focal plane in the imaged stack, thus ensuring that motion is minimized at each axial position. An alternative approach is to image only high-SNR images and generate corresponding low-SNR training data semi-synthetically, by adding artificial noise to high SNR data from living samples [11,30].

If collecting high SNR ground truth images is impractical or impossible, one can use unsupervised deep learning methods (**Fig. 1b**), which do not require high SNR "ground truth" images during training. Instead, these methods assume spatial and/or temporal sample continuity and operate directly on a body of noisy images [32,118,124]. Another advantage of unsupervised methods is that they can be trained on and applied to the same body of data, removing the possibility that the subcellular structures the network was trained on are different from the ones in the data to be denoised. This circumvents a potential problem associated with supervised methods, which may generalize poorly to unseen or underrepresented structures in the training data.

Each of these strategies may be worth testing, which is increasingly feasible since many methods are open-source and offer tutorials. Nevertheless, trying multiple approaches entails considerable effort, and training a network can take hours – often longer than it takes to acquire the data. Moreover, the expertise may not be accessible to every microscopist or biologist. Generally, supervised training with high-quality data leads to the best denoising and image restoration performance (**Fig. 2**). Still, unsupervised denoising may be all that is needed to enable the desired analyses and is thus worth considering as a first step.

Perhaps the greatest caveat when using any deep learning method is that the network output is only a prediction of the 'ground truth', and thus caution should be exercised in interpreting such output (**Fig. 1d**). Furthermore, trained networks must themselves deal with uncertainties, and do so in different ways. Recent denoising approaches can sample multiple denoised "interpretations" of raw input data. These samples are drawn from a previously learned distribution of reasonable data appearances. Most of the above-mentioned supervised and unsupervised approaches, however, return single outputs, which are generally closer to the "average" of all possible denoised interpretations. While these details are known to method developers, they are often not appreciated by users. This underscores the importance of an open discourse and consistent training efforts in this area.

Several related points are worth remembering in the specific context of denoising. First, any denoising method's performance will decrease in the presence of increasing noise (**Fig. 2**). It can be useful to assess the SNR, illumination intensity, or exposure at which the noise is 'too much', and the prediction quality becomes unacceptable, remembering that this will be sample and structure dependent [29]. Obtaining high SNR images as 'sanity checks' is useful if biological interpretation depends critically on the quality of the denoised output. Second, the degree to which the prediction matches the ground truth depends on the structure, with more faithful restoration of larger structures, and those with higher label density and input SNR. Third, networks tend to generalize poorly when presented with images dissimilar to the data they are trained on (**Fig. 1b, c**). If quantitative conclusions based on signal intensities are desirable, we advise checking if network predictions preserve linearity [29], e.g., by comparing ROI intensities in various locations throughout the raw and restored image data and plotting them against each other [125].

**Box 4: Practical considerations when using computation to improve spatial resolution**

When using deconvolution, one caveat is that algorithms generally assume the PSF is spatially invariant, meaning a single blurring function accurately describes the blurring at all locations in the image volume. However, in most microscopes and for most samples, the blurring is spatially variant, particularly as a function of depth due to optical aberrations, or in specialized cases, where the PSF varies by design across the field of view [126]. Accounting for this variation can lead to better deconvolution but is computationally taxing [127], although a recent DL-based implementation [128] describes an orders-of-magnitude improvement in speed. Another practicality to keep in mind is that the iterative nature of deconvolution can make even the relatively simple RL algorithm time consuming to execute, so that processing times significantly outweigh data acquisition times, particularly for large datasets. Deciding when to stop the iterative deconvolution process is another challenge, with more iterations improving the image (at the expense of additional computational overhead), but too many iterations leading to noise amplification or unrealistic 'oversharpening' of structures (**Fig. 3 a-c**). Variants of RL or deep learning approaches designed to mimic deconvolution can considerably shorten this computational overhead or bypass the stopping criterion [59,129].

When using deep learning to deblur data, as for denoising applications (**Box 3**), a major requirement which can limit the use of these methods is acquiring high quality training data. Acquiring pairs of low- and high-resolution image pairs on the same microscope is preferred, since it allows easier and more accurate registration between pairs, but may be difficult in practice. In such difficult cases, high-resolution data can be computationally degraded to yield semi-synthetic low-resolution data, with high accuracy if the PSF and noise of the latter can be estimated [11,29,30]. In some cases of especially simple image content, the network may be trained on entirely synthetic ground truth that has been blurred and degraded to resemble real biological structures; this approach was used to deblur images of microtubules and secretory granules in living cells [11].

A key challenge is assessing the degree to which the deep learning prediction can be trusted. Simulations and experiments indicate that predictions degrade in the presence of increasing noise in the input data, and as the degree of resolution enhancement increases – with increasingly obvious deterioration for networks trained to achieve more than 2-fold resolution enhancement over the raw data [29,32,125] (**Fig. 3d, e**). Perhaps the most prudent advice is – as for any imaging experiment – to perform controls, ideally validating the degree of resolution enhancement with other methods, such as traditional super-resolution microscopy.

In the future, better reference datasets should enable more comprehensive comparisons across different tools, making it easier for biologists to determine which of the rapidly growing zoo of methods can best address a given problem. Still, it is important that users have realistic expectations about what AI models can, and cannot, predict, so their work can remain rigorous and reproducible. For example, AI cannot fully recover fine details of structures in diffraction limited microscopy, since information at those spatial scales is lost during image formation. At best, AI can make predictions about what such structures might look like, based on the raw input image and the content-aware data prior learned from the training data [11]. The more stereotyped the structure is, and the more similar it is to the training data, the better the predictions are likely to be (**Fig. 1c**).